\documentclass[prc,aps,fleqn,twocolumn,twoside]{revtex4}

\usepackage{latexsym,exscale}
\usepackage{epsfig}
\usepackage{amsmath,amssymb,amsfonts}

\begin{document}

\title{Hadron production in heavy ion collisions: \\ Fragmentation and 
   recombination from a dense parton phase}
\author{R.~J.~Fries}
\affiliation{Department of Physics, Duke University, Durham, NC 27708}
\author{S.~A.~Bass}
\affiliation{Department of Physics, Duke University, Durham, NC 27708}
\affiliation{RIKEN BNL Research Center, Brookhaven National Laboratory, 
        Upton, NY 11973, USA}
\author{B.~M\"uller}
\affiliation{Department of Physics, Duke University, Durham, NC 27708}
\author{C.~Nonaka}
\affiliation{Department of Physics, Duke University, Durham, NC 27708}
\date{\today}

\begin{abstract}
We discuss hadron production in heavy ion collisions at RHIC. We argue
that hadrons at transverse momenta $P_T < 5$ GeV are formed by
recombination of partons from the dense parton phase created in central 
collisions at RHIC. We provide a theoretical description of the recombination
process for $P_T > 2$ GeV. Below $P_T = 2$ GeV our results smoothly match
a purely statistical description.
At high transverse momentum hadron production is well described in the 
language of perturbative QCD by the fragmentation of partons. 
We give numerical results for a variety of hadron spectra, ratios and nuclear
suppression factors. 
We also discuss the anisotropic flow $v_2$ and give results based
on a flow in the parton phase.
Our results are consistent with the existence of a parton phase at RHIC
hadronizing at a temperature of 175 MeV and a radial flow velocity of 0.55$c$.
\end{abstract}

\preprint{DUKE-TH???}

\maketitle

\section{Introduction}

Recent data from the Relativistic Heavy Ion Collider (RHIC) have shown a 
strong nuclear suppression of the pion yield at transverse momenta larger than
2 GeV/$c$ in central Au + Au collisions, compared to $p+p$ interactions 
\cite{PHENIX}. This is widely seen as the experimental confirmation of
jet quenching, the phenomenon that high energy partons lose energy when 
they travel through the hot medium created in a heavy ion collision
\cite{GyulWang:94,BDMS:01,Muller:02}, entailing a suppression of intermediate 
and high $P_T$ hadrons. 

However, the experiments at RHIC have provided new puzzles. The amount of 
suppression seems to depend on the hadron species. In fact, in the production 
of protons and antiprotons between 2 and 4 GeV/$c$ the suppression seems to 
be completely absent. Generally, pions and kaons appear to suffer from a 
strong energy loss while baryons and antibaryons do not. Two stunning 
experimental facts exemplify this \cite{PHENIX-B,STAR-B,STAR-L,PHENIX-L}.
First, the ratio of protons over positively charged pions is equal or above 
one for $P_T > 1.5\text{ GeV}/c$ and is approximately constant up to 4 GeV/$c$.
Second, the nuclear suppression factor $R_{AA}$ below 4 GeV/$c$ is close to 
one for protons and lambdas, while it is about 0.3 for pions.

There have been recent attempts to describe the different behavior of baryons
and mesons through the existence of gluon junctions \cite{GyulVit:01} 
or alternatively through 
recombination as the dominant mechanism of hadronization 
\cite{FMNB:03,GreKoLe:03}.
The recombination picture has attracted additional attention due to the
observation that the elliptic flow pattern of different hadron species can 
be explained by a simple recombination mechanism 
\cite{LinKo:02,Voloshin:02,MoVo:03,LinMol:03}. 
The anisotropies $v_2$ for the different hadrons are compatible with a 
universal value of $v_2$ in the parton phase, related to the hadronic flow  
by factors of two and three depending on the number of valence quarks 
\cite{Sorensen:03}.

In this work we elaborate on the arguments presented in \cite{FMNB:03}. 
We will present the formalism as well as new numerical results.
The competition between recombination and fragmentation
delays the onset of the perturbative/fragmentation regime to relatively high 
transverse momentum of 4--6 GeV/$c$, depending on the hadron species.
This is the explanation for several key observations at RHIC:
\begin{itemize}
\item the two component form of hadron transverse momentum spectra, including 
   an exponential part and a power law tail with a transition between 
   4 and 6 GeV/$c$.
\item the very different behavior of the nuclear suppression factors 
   $R_{AA}$ of mesons and baryons.
\item the particle dependence of the elliptic flow.
\item the unusually large baryon/meson ratios.
%\item the participant scaling observed by PHOBOS \cite{Roland:02} up to
%   5 GeV/$c$.
\end{itemize}
In addition, our calculation clarifies the range of applicability 
of perturbative calculations including energy loss.

The paper is organized as follows. In the next section we explain 
why fragmentation might not be the dominant mechanism of hadronization 
at intermediate transverse momenta of a few GeV/$c$ in heavy ion collisions.
We discuss the fundamental principles of recombination and fragmentation.
In Sec.\ III we present the theoretical framework of recombination
in more detail and also discuss its shortcomings. In particular, we address
the question of applicability at low transverse momentum. In Sec.\ IV we
introduce our parametrization of the parton spectrum and discuss further 
calculational details, and in Sec.\ V we present numerical results on spectra, 
hadron ratios, nuclear suppression and elliptic flow. Sec.\ VI summarizes 
our work.

\section{Fragmentation vs.\ Recombination}

\subsection{Fragmentation of partons}

Inclusive hadron production at sufficiently large momentum transfer 
can be described by perturbative quantum chromodynamics (pQCD). 
The invariant cross section for
a hadron $h$ with momentum $P$ can be given in factorized form
\cite{Owens:86}
\begin{equation}
  \label{eq:fracmaster}
  E \frac{d \sigma_h}{d^3 P} = \sum_a \int\limits_0^1 \frac{d z}{z^2} 
  D_{a\to h}(z) E_a \frac{d \sigma_a}{d^3 P_a}.
\end{equation}
The sum runs over all parton species $a$ and $\sigma_a$ is the cross section
for the production of parton $a$ with momentum $P_a = P/z$.
Thus the parton production cross section has to be convoluted with
the probability that parton $a$ fragments into hadron $h$. The probabilities
$D_{a\to h}(z)$ are called fragmentation functions \cite{CoSo:81}. Like 
parton distributions they are non-perturbative quantities. However they
are universal and once measured, e.g.\ in $e^+ e^-$ annihilations, they can
be used to describe hadron production in other hard QCD processes.

Using (\ref{eq:fracmaster}) we can estimate the ratio of protons and pions. 
Taking, e.g., the common parametrization of Kniehl, Kramer
and P\"otter (KKP) \cite{KKP:00}, the ratio $D_{a\to p}/ D_{a\to \pi^0}$
is always smaller than 0.2 for each parton $a$. This reflects the well known
experimental fact that pions are much more abundant than protons in the
domain where pQCD is applicable. The excess of pions over protons even holds 
down to very low $P_T$, smaller than 1 GeV, where perturbative calculations 
are no longer reliable. In that domain one can argue that the difference in 
mass, $M_p \gg M_\pi$, lays a huge penalty on proton production. The small
value of the $p/\pi^0$ ratio predicted by these calculations over
the entire range of $P_T$ is the 
reason why the ratio $p/\pi^0 \sim 1$ measured at RHIC is so surprising.

It has been suggested that the fragmentation functions $D_{a\to h}(z)$ can be 
altered by the environment \cite{GW:00,WaWa:02}. 
The energy loss of the propagating 
parton in the surrounding medium leads, in first approximation, to a rescaling 
of the variable $z$. This would affect all produced hadrons in the same way, 
and thus cannot explain the observations at RHIC. In a picture
with perturbative hadron production and jet quenching alone, the different
behavior of hadrons can not be described by one consistent sets of energy loss
parameters. To save the validity of the purely perturbative approach species 
dependent non-perturbative contributions to the fragmentation functions
have to be introduced {\it ad hoc} to explain the data \cite{Wang:03}.

Perturbative hadron production consists of three steps: production
of a parton in a hard scattering, propagation and interaction with a medium,
and finally hadronization of the parton. 
Only modifications in our understanding of hadronization are able to provide 
an explanation of the experimental observations, since the other steps
are blind to the hadron species that will eventually be created.

\subsection{From fragmentation to recombination}

For the production of a hadron with momentum $P$ via fragmentation we need to 
start with a parton with momentum $P/z > P$. The fragmentation functions
favor very small values of $z$, {\it i.e.}\ the situation where the energy of
the fragmenting parton is not concentrated in one hadron. 
On the other hand, the transverse momentum spectrum of partons is steeply 
falling with $P_T$. This makes it clear that fragmentation is a rather
inefficient mechanism for the production of high $P_T$ hadrons, since
it has to overcome the limited
availability of partons at even higher transverse momentum.
As a result, the average $\langle z \rangle$ is larger than what is
expected from the shape of the fragmentation functions. For 
pion production, $\langle z \rangle$ is about 0.6 for the production from a 
valence quark, 0.4 for a sea quark and 0.5 for a gluon in the range 
$2\text{ GeV}/c \le P_T \le 10\text{ GeV}/c$ for leading order KKP 
fragmentation functions.

An outgoing high energy parton is not a color singlet and will therefore
have a color string attached. The breaking of the string will initiate
the creation of quark antiquark pairs until there is a entire jet of partons, 
which have to share the energy of the initial parton. They will finally
turn into many hadrons. The creation of several 
hadrons from one fragmenting parton is the reason why fragmentation functions 
prefer small values of $z$. If phase 
space is already filled with partons, a single parton description might
not be valid anymore. Instead one would have to introduce 
higher twist (multiple
parton) fragmentation functions. In the most extreme case, if
partons are abundant in phase space, they might simply recombine into hadrons.
This means that a $u$ and a $\bar d$ quark that are ``close'' to each other
in phase space can bind together to form a $\pi^+$. The scale of being close
will be set by the width of the pion wave function. In this scenario the
total pion momentum will be just the sum of the individual quark momenta.
We immediately notice that this recombination mechanism is very 
efficient for steeply falling
spectra: in order to produce a 5 GeV pion we can start with two quarks 
having (on average) about 2.5 GeV/$c$ transverse momentum and each being 
therefore far more abundant (on average) than a 10 GeV/$c$ parton that could 
produce the pion via fragmentation. Of course the recombining 
partons must be close in phase space, i.e. recombination will be suppressed
if the phase space density is low.

Recombination can be interpreted as the most ``exclusive'' 
form of hadronization, the endpoint of a hypothetical resummation of
fragmentation processes to arbitrary twist. We cannot achieve a quantitative 
understanding of this at the moment. However, we can try to find an 
effective description which can be tested against observable consequences.
In this work we will advocate a simple model for recombination and compare
it with single parton fragmentation. These two mechanisms of hadron
production compete differently depending on the phase space density of 
partons. From the above we understand that the competition between 
fragmentation and recombination is dominated by the slope and the 
absolute value of the phase space distribution of partons. Below we show 
that recombination always wins over fragmentation for an exponentially 
falling parton spectrum, but that fragmentation takes over if the 
spectrum has the form of a power law, as it is provided by pQCD. We will 
apply this insight to hadron production in relativistic heavy ion collisions 
at midrapidity and transverse momenta of a few GeV/$c$ where we expect a 
densely populated phase space. For the recombination of three quarks into a 
proton the momenta of three partons have to be added up, but only two momenta 
in the case of a pion. 
Assuming an exponential parton spectrum this implies for a proton a 
distribution $\sim [\exp(-P_T/3)]^3$ and for pions $\sim [\exp(-P_T/2)]^2$, 
predicting a constant $p/\pi^+$ ratio where the value is determined by simple 
counting of quantum numbers \cite{FMNB:03}. We will show that some of the 
surprising experimental results from RHIC can be explained in this way.

\subsection{The recombination concept}

The idea of quark recombination was proposed long ago to describe 
hadron production in the forward region of $p+p$ collisions
\cite{DasHwa:77}. This was later justified by the discovery of the
leading particle effect, the phenomenon that, in the forward direction 
of a beam of hadrons colliding with a target, the production of hadrons 
sharing valence quarks with the beam hadrons are favored. E.g.\ in the Fermilab
E791 fixed target experiment with a 500 GeV $\pi^-$ beam \cite{E791:96} the 
asymmetry 
\begin{equation}
  \alpha(x_F) = \frac{d\sigma_{D^-}/d x_F - d\sigma_{D^+}/d x_F}{
  d\sigma_{D^-}/d x_F + d\sigma_{D^+}/d x_F}
\end{equation}
between $D^-$ and $D^+$ mesons 
grows nearly to unity when the Feynman variable $x_F$, measuring the 
longitudinal momentum relative to the beam momentum, approaches 1. 
Fragmentation would predict
this asymmetry to be very close to zero. However, recombination of the 
$\bar c$ quark from a $c\bar c$ pair produced in a hard interaction
with a $d$ valence quark from the $\pi^-$, propagating in forward direction 
with large momentum, is highly favored compared to the recombination of the 
$c$ with a $\bar d$ which is only a sea quark of the $\pi^-$. This leads to
the enhancement of $D^-$ over $D^+$ mesons in the forward region.

The leading particle effect is a clear signature for the existence of
recombination as a hadronization mechanism and has been addressed in several
publications recently \cite{AnMaHe:01,BraJiaMe:02}.
In this case recombination is favored over fragmentation only in a certain 
kinematic situation (the very forward direction), which is a only a small 
fraction of phase space. 

In central heavy ion collisions many more partons are produced than in 
collisions of single hadrons. The idea that recombination may then be 
important for a wide range of rapidities -- and 
at least up to moderate transverse momenta -- 
was advocated before \cite{recomb1,recomb2,recomb3}. However, it was only 
recently that RHIC data indicated that recombination
could indeed be a valid approach up to 
surprisingly high transverse momenta of a few GeV/$c$. 

Charm and heavy hadron production have the advantage that the heavy quark mass
provides a large scale that permits a more rigorous treatment of the 
recombination process \cite{BraJiaMe:02}. The description of recombination 
into pions and protons seems to be theoretically less rigorous. However, 
a simple 
counting of quantum numbers in a picture where the structure of hadrons 
is dominated by their valence quarks often provides surprisingly good results. 
This has been pointed out for particle spectra and ratios 
\cite{HwaYa:02,FMNB:03,GreKoLe:03b} and for elliptic flow 
\cite{LinKo:02,Voloshin:02,MoVo:03}. Most of the work so far has stayed 
on a qualitative level without quantitative predictions. Recombination of 
$D$ mesons in heavy ion collisions has been investigated in \cite{RappShu:03}.

We will argue below that the counting of quantum numbers is a good 
description of the recombination process for intermediate momenta. 
The formalism will set the stage to obtain quantitative results in this
regime by recombining quarks from a possibly thermalized phase. 
We know that the parton phase will not behave like a perturbative plasma
near the hadronization point \cite{KaKaLaLu:99}. Instead quarks  at 
hadronization will be effective degrees of freedom exhibiting a mass and
gluons will disappear as dynamical degrees of freedom.
We will assume here that the effective quarks behave like constituent quarks 
and that there are no dynamical gluons.

This is different from the work of Greco {\it et al.} \cite{GreKoLe:03} who 
suggested to recombine one perturbative quark with thermal quarks, leading 
to an additional contribution at the transition region between the pure 
thermal phase dominating below 5 GeV/$c$ and the pure fragmentation regime 
dominating above 5 GeV/$c$. A similar form of this pick-up reaction of 
a perturbative parton was recently proposed in the context of the sphaleron
model \cite{SolShu:03}.

One can also attempt to extend the recombination concept to low $P_T$, 
however, the theoretical situation is much more ambiguous there. The main 
reasons are that the simple counting of quantum numbers violates energy and 
entropy conservation at low $P_T$, where the bulk of the hadron yield resides.
Nevertheless, once recombination is recognized to be the dominant hadronization
mechanism at intermediate $P_T$ from 2 to 5 GeV/$c$, it is quite reasonable to 
expect that this mechanism extends down to very small transverse momentum, 
where quarks are even more abundant. However the theoretical description will 
only be on solid ground once the problems of energy and entropy conservation
are addressed properly.

\subsection{A non-relativistic model}
\label{sec:simple}

For a first estimate, let us consider a simple static model for the 
recombination process. We start with a system of
quarks and antiquarks which are 
homogeneously distributed in a fixed three-dimensional 
volume $V$ with phase space distributions $w(\bf{p})$, so that the number of 
quarks or antiquarks with a particular set $a$ of quantum numbers 
(color, spin, isospin) is
\begin{equation}
  N_a = V \int \frac{d^3 p}{(2\pi)^3} w_a({\bf p})
\end{equation}
Hadronization is assumed to occur instantaneously throughout the volume. 
The distributions $w({\bf p})$ are supposed not to change during the 
hadronization process, {\it i.e.}\ the quarks are assumed to be quasi-free.

%\subsubsection{Mesons}

The spatial wave functions for a two particle
quark/antiquark state with momenta ${\bf p}_1$ and ${\bf p}_2$ and a meson 
bound state with momentum $\bf P$ are
\begin{align}
  \langle x \, | \, q,{\bf p}_1{\bf p}_2 \rangle &= V^{-1} 
  e^{i ({\bf p}_1{\bf x}_1+{\bf p}_2{\bf x}_2}) \\
  \langle x \, | \, M,{\bf P} \rangle &= V^{-1/2} e^{i{\bf P \cdot R}} 
  \, \varphi_M({\bf y})
\end{align}
respectively, with the center of mass and relative coordinates 
${\bf R} = ({\bf x}_1 + {\bf x}_2)/2$ and ${\bf y}= {\bf x}_1 - {\bf x}_2$
for the two quarks in the meson system. To keep our notation simple we
omit the proper antisymmetrization of multi fermion states, since all
combinatorial factors will cancel in the final result. 
The internal meson wave functions is normalized as
\begin{equation}
  \label{eq:pinorm}
  \int d^3 y \> | \varphi_M({\bf y}) |^2 = 1.
\end{equation}
 
The overlap amplitude is given by
\begin{multline}
  \langle q, {\bf p}_1{\bf p}_2 \, | \pi, {\bf P} \rangle  \\
%  = V^{-3/2} \int d^3 x_1 d^3 x_2 \, 
%  e^{-i ({\bf p}_1{\bf x}_1+{\bf p}_2{\bf x}_2}) e^{i{\bf P R}} \Phi({y}) \\
  = \frac{(2\pi)^3}{V^{3/2}} \delta^3 ({\bf P} - {\bf p}_1 - {\bf p}_2) 
  \hat \varphi_M({\bf q}).
\end{multline}
Here, we have introduced the relative momentum 
${\bf q}=({\bf p}_1-{\bf p}_2)/2$, 
conjugate to ${\bf y}$, and $\hat\varphi_M({\bf q})$ is the Fourier
transformed wave function.
%\begin{equation}
%  \tilde\varphi_M(q) = \int d^3 r \, e^{-i {\bf q r}} \varphi_M(r) = N_\pi 
%  e^{-q^2 /2 \Lambda_\pi^2}
%\end{equation}
%is the Fourier transformed wave function.
The squared amplitude is
\begin{equation}
  | \langle q, {\bf p}_1{\bf p}_1 \, | M, {\bf P} \rangle |^2 =
  \frac{(2\pi)^3}{V^2} \delta^3 ({\bf P}-{\bf p}_1-{\bf p}_2) 
  \big|\hat\varphi_M({\bf q})\big|^2 .
\end{equation}

We conclude that the total number of pions found in the quark/antiquark 
distribution is 
\begin{multline} 
  N_M = C_M V^3 \int\frac{d^3 P}{(2\pi)^3}\frac{d^3 p_1}{(2\pi)^3}
  \frac{d^3 p_2}{(2\pi)^3} \> \\
  \times w({\bf p}_1)w({\bf p}_2) \>
  \big| \langle q, {\bf p}_1{\bf p}_1 \, | M, {\bf P} \rangle \big|^2 
\end{multline}
with a degeneracy factor $C_M$.
The momentum distribution of the pions is given by
\begin{multline}
  \label{eq:pionres1}
  \frac{d N_M}{d^3 P} = 
  C_\pi \frac{V}{(2\pi)^3} \int \frac{d^3 q}{(2\pi)^3} \\  \times
  w\Big(\frac{\bf P}{2} +{\bf q} \Big) 
  w\Big(\frac{\bf P}{2} -{\bf q} \Big) 
  \big|\hat\varphi_M({\bf q})\big|^2 
\end{multline}

From above equation the spectra of mesons can be calculated for given
quark distributions. This will require knowledge of $\varphi_M$. 
The wave function has some width $\Lambda_M$, and we assume that it
drops rapidly for $|{\bf q}| > \Lambda_M$.
Let us study the kinematic region where $P\gg \Lambda_M$. The integral over 
the relative momentum ${\bf q}$ is dominated by
values $|{\bf q}| \sim \Lambda_M$ and thus we can assume that 
$|{\bf q}| \ll |{\bf P}|$.
We apply a Taylor expansion 
\begin{widetext}
\begin{equation}
  w\Big(\frac{\bf P}{2} +{\bf q} \Big) w\Big(\frac{\bf P}{2} -{\bf q} \Big) 
  = w\Big(\frac{\bf P}{2} \Big)^2 + \sum_{ij} q_i q_j \left[
  w\Big(\frac{\bf P}{2} \Big) \partial_i \partial_j 
  w\Big(\frac{\bf P}{2} \Big) - \partial_i w\Big(\frac{\bf P}{2} \Big) 
  \partial_j w\Big(\frac{\bf P}{2}\Big) \right] + {\mathcal O} (q^3),
\end{equation}
\end{widetext}
where the first order term in the expansion vanishes.

From the lowest order term in the expansion we get a contribution to the
meson spectrum which is independent of the shape of the wave function. Only
the normalization (\ref{eq:pinorm}) enters and leads to a universal term
\begin{equation}
  \label{eq:pionleading}
  C_\pi \frac{V}{(2\pi)^3} w({\bf P}/{2})^2.
\end{equation}
The second order term (like all higher order terms) in the expansion generates
a correction depending on the shape of the wave function. 
For the sake of simplicity, let us assume a Gaussian shape
\begin{equation}
  \hat\varphi_M({\bf q}) = \mathcal{N}_M   e^{-{\bf q}^2 / 2\Lambda_M^2}. 
\end{equation}
The normalization factor 
is determined by (\ref{eq:pinorm}) as 
$\mathcal{N}_M^2 = (2 \sqrt{\pi}/\Lambda_M)^3$.
For this example we have the second order correction
\begin{equation}
  \label{eq:pioncorr}
  C_\pi \frac{V}{(2\pi)^3} \frac{\Lambda_M^2}{2} 
  \left[ w({\bf P}/{2}) \triangle w({\bf P}/{2}) - \big(\nabla w({\bf P}/{2})
  \big)^2 \right].
\end{equation}

We want to emphasize once more that in the limit $P \gg \Lambda_M$ 
the exact shape of the wave function is not important. In a relativistic 
framework this statement is softened
by the fact that we consider the hadron formation in a boosted frame,
where $P$ is large. Therefore $\Lambda_M$, which is of order 
$\Lambda_{\text{QCD}}$ in the rest frame of the hadron, will be dilated.

As an example, let us assume an exponential parton distribution 
of the form $w({\bf p})= 
e^{-p/T}$:
The meson spectrum at large $P$ is then given by
\begin{equation}
  \frac{d N_\pi}{d^3 P} = C_\pi \frac{V}{(2\pi)^3} e^{-P/T}
  \left[ 1 -  \frac{2  \Lambda_M^2}{T P}  \right],
\end{equation}
using (\ref{eq:pionleading}),(\ref{eq:pioncorr}).
%Apparently we are neglecting the masses of the particles here, but this is
%consistent with the assumption of a large $P$. 
The second order term introduces a power correction of order 
$\Lambda_M^2 / TP$ to the universal result, depending both on the width of 
the wave function and the slope of the parton distribution.

%\subsubsection{Baryons}

For nucleons we start with three quarks at coordinates ${\bf x}_i$.
We introduce center of mass and relative coordinates ${\bf R} = ({\bf x}_1 +
{\bf x}_2 + {\bf x}_3 )/3$, ${\bf y} = ({\bf x}_1 + {\bf x}_2)/2 - 
{\bf x}_3 $ and ${\bf z} = {\bf x}_1 - {\bf x}_2 $.
%Again we assume a simple Gaussian wave function for the nucleon
%\begin{equation}
%  \Psi({\bf y},{\bf z})= \Psi(y,z) = \mathcal{N}_B e^{-y^2 \Lambda_{B1}^2 /2}
%  e^{-z^2 \Lambda_{B2}^2 /2}.
%\end{equation}

The overlap amplitude between a 3 quark state and a baryon with momentum ${\bf
P}$ is
\begin{equation}
  \langle q, {\bf p}_1{\bf p}_2{\bf p}_3 \, | B, {\bf P} \rangle 
  = \frac{(2\pi)^3}{V^{2}} \delta^3 ({\bf P} -{\bf p}_1-{\bf p}_2-{\bf p}_3) 
  \hat\varphi_B({\bf q},{\bf s}).
\end{equation}
Here $\hat\varphi_B({\bf q},{\bf s})$ is the baryon wave function in momentum 
space, depending on the relative momenta ${\bf q}$ and ${\bf s}$ conjugate
to ${\bf y}$ and ${\bf z}$ respectively.

Hence the baryon distribution is given by
\begin{multline}  
  \label{eq:nuclres1}
  \frac{d N_B}{d^3 P} = 
  C_B \frac{V}{(2\pi)^3} \int \frac{d^3 q}{(2\pi)^3} \frac{d^3 s}{(2\pi)^3} \> 
  \big|\hat\varphi_B({\bf q},{\bf s})\big|^2
  \\ \times
  w\Big(\frac{\bf P}{3} +\frac{\bf q}{2} +{\bf s} \Big) 
  w\Big(\frac{\bf P}{3} +\frac{\bf q}{2} -{\bf s} \Big)
  w\Big(\frac{\bf P}{3} - {\bf q}\Big) .
\end{multline}
$C_B$ is the appropriate degeneracy factor.
In the region where the nucleon momentum ${\bf P}$ is large compared to the
intrinsic width of the wave function, we can again expand the
product of the quark distribution functions.
The leading term is universal and contributes
 \begin{equation}
  \label{eq:nuclleading} 
  C_\pi \frac{V}{(2\pi)^3} w({\bf P}/{3})^3
\end{equation}
to the nucleon distribution.
The second order term provides a correction
\begin{multline}
  \label{eq:nuclcorr}
  C_\pi \frac{V}{(2\pi)^3} \frac{1}{2} \left( \Lambda_{B2}^2  + \frac{3}{4}
  \Lambda_{B1}^2 \right) \\
  \left[ w({\bf P}/{3})^2 \triangle w({\bf P}/{3}) - 
  w({\bf P}/{3})\big(\nabla w({\bf P}/{3}) \big)^2 \right],
\end{multline}
assuming a normalized Gaussian wave function
\begin{equation}
  \hat\varphi_B({\bf q},{\bf s}) = \mathcal{N}_B 
  e^{-{\bf q}^2 /2 \Lambda_{B1}^2} e^{-{\bf s}^2 /2 \Lambda_{B2}^2}.
\end{equation}

\subsection{First conclusions}

Let us summarize and analyze our first results. The transverse spectrum of
mesons from recombination is proportional to $C_M w^2(P_T/2)$ whereas 
fragmentation would generate a distribution proportional to 
$D(z) \otimes w(P_T/z)$. For an exponential parton spectrum 
$w=e^{-P_T/T}$ the ratio of recombination to fragmentation is
\begin{equation}
  \frac{R}{F} = \frac{C_M}{\langle D \rangle} 
  e^{-\frac{P_T}{T} \left( 1- \frac{1}{\langle z \rangle} \right) }
\end{equation}
where $\langle D \rangle$ and $\langle z \rangle <1 $ are average values
of the fragmentation function and the scaling variable. Obviously, for
large $P_T$ one always gets $R/F > 1$. In other words, recombination 
always wins over fragmentation at ``high'' $P_T$ for an exponential 
parton spectrum. The same is true for baryons as well as mesons.

Now let us consider a parton distribution given by 
a power law spectrum $w=A (P_T/\mu)^{-\alpha}$ with
a scale $\mu$ and $\alpha >0 $. Then the ratio of recombination over 
fragmentation is
\begin{equation}
  \frac{R}{F} = \frac{C_M A}{\langle D \rangle} 
  \left(\frac{4}{\langle z \rangle} \right)^\alpha 
  \left(\frac{P_T}{Q}\right)^{-\alpha}
\end{equation}
and fragmentation ultimately dominates at high $P_T$. Again, this holds
both for mesons and baryons. This implies that fragmentation from an 
exponential spectrum and recombination from a power law spectrum are 
suppressed. We will thus omit these contributions in our work. 

One might ask whether these considerations are still valid in the case
of more realistic formulation of recombination. It turns out that the
basic formulae obtained above are still valid in a relativistic 
description. Deviations are less than 20\% for $P_T > 2$ GeV/$c$.

Given an exponential parton spectrum we note that recombination
yields a a constant baryon-to-meson ratio.
The ratio is then only determined by the degeneracy factors
\begin{equation}
  \frac{dN^R_B}{dN^R_M} = \frac{C_B}{C_M}.
\end{equation}
Just counting the hadron degeneracies, the direct $p/\pi^0$ ratio (neglecting 
protons and pions from secondary hadronic decays) would be $\sim 2$, in 
contrast to $\sim 0.2$ from fragmentation in pQCD. We will later see that 
finite mass effects and superposition with the fragmentation process will 
bring down the $p/\pi^0$ ratio from 2 to approximately 1 in the range between 
2 and 4 GeV/$c$ transverse momentum.

\section{The recombination formalism}

In this section, we turn to a better description of recombination. 
This will require a more realistic model of the parton
phase including longitudinal and transverse expansion as well as
an improved space-time picture.

Let us consider a system of quarks and antiquarks evolving in Minkowski space.
We choose a spacelike hypersurface $\Sigma$ on which recombination of these 
partons into hadrons occurs. In the simplest scenario, that could be just a
slice of Minkowski space with fixed time $t_0$, leading  back to the case
we described in the previous subsection.

It has been discussed in the literature \cite{WieToHei:98,ScheiHei:99}, how 
the freeze-out can be smeared around the hypersurface to account for 
a finite hadronization time. However, RHIC experiments suggest a very
rapid freeze-out. The measured two-particle correlation functions are 
consistent with an extremely short emission time in the local rest frame, 
suggesting a sudden transition after which the individual hadrons interact 
only rarely \cite{RHIC-HBT}.

This can be understood by the fact that the hadronization time in the 
laboratory frame is Lorentz contracted. The Lorentz factors $\gamma_T$ 
are 3.35 for a 3 GeV/$c$ proton and 22 for a pion of the same $P_T$.
The hadronization time --- even if it is more than 1 fm/$c$ in the rest 
frame of the hadron --- will be experienced as considerably shorter by fast
hadronizing particles.

\subsection{Wigner function formalism}

Recombination has already been considered before in a covariant form
utilizing Wigner functions for the process of baryons coalescing into light 
nuclei and clusters in nuclear collisions \cite{DHSZ:91,ScheiHei:99}. 
Here, we will provide a derivation for the recombination of a quark 
antiquark pair
into a meson. The generalization to a three quark system recombining into a
baryon is straightforward.

By introducing the density matrix $\hat \rho$ for the system of partons, 
the number of quark-antiquark states that we will interpret as mesons is given 
by
\begin{equation}
  \label{eq:pinumb}
  N_M = \sum_{ab}\int \frac{d^3 P}{(2\pi)^3} \> \langle M ;{\bf P} | \> \hat 
  \rho_{ab} \> | M ;{\bf P} \rangle
\end{equation}
Here $| M ;{\bf P} \rangle$ is a meson state with momentum ${\bf P}$ and the
sum is over all combinations of quantum numbers --- flavor, helicity and 
color --- of valence partons that contribute to the given meson $M$.
We insert complete sets of coordinates
\begin{multline}
  N_M = \int \frac{d^3 P}{(2\pi)^3} d^3 \hat r_1 d^3 \hat r'_1 d^3 
  \hat r_2 d^3 \hat r'_2
  \> \langle M;{\bf P} |  \> \hat r_1 , \hat r_2 \rangle \\  \times
  \langle \hat r_1 , \hat r_2 \> | \> \hat \rho \> | \> 
  \hat r'_1 , \hat r'_2  \rangle \langle \hat r'_1 , \hat r'_2 | M ; 
  {\bf P} \rangle.
\end{multline}
and change the variables to ${\bf r}_{1,2} = (\hat {\bf r}_{1,2} + \hat 
{\bf r}'_{1,2})/2$ and ${\bf r}'_{1,2} = \hat {\bf r}_{1,2} - 
\hat {\bf r}'_{1,2}$.
We define the 2-parton Wigner function 
$W_{ab}({\bf r}_1,{\bf r}_2;{\bf p}_1,{\bf p}_2)$ as
\begin{widetext}
\begin{equation}
  \bigg\langle {\bf r}_1 - \frac{{\bf r}'_1}{2} , {\bf r}_2 - 
  \frac{{\bf r}'_2}{2} \> \bigg| 
  \> \hat \rho \> \bigg| {\bf r}_1 + \frac{{\bf r}'_1}{2} , {\bf r}_2 +
  \frac{{\bf r}'_2}{2} \bigg\rangle 
  = \int \frac{d^3 p_1}{(2\pi)^3}\frac{d^3 p_2}{(2\pi)^3}
  e^{-i {\bf p}_1\cdot {\bf r}'_1} e^{- i {\bf p}_2 \cdot {\bf r}'_2} 
  W_{ab}({\bf r}_1,{\bf r}_2;{\bf p}_1,{\bf p}_2)
\end{equation}
\end{widetext}
and introduce a notation for the meson wave function $\varphi_M$
\begin{equation}
  \bigg\langle {\bf r}_1 + \frac{{\bf r}'_1}{2} , 
  {\bf r}_2 + \frac{{\bf r}'_2}{2} \> \bigg| \> \pi ; {\bf P} \bigg\rangle =
  e^{- i {\bf P} \cdot ({\bf R} + {\bf R}'/2)} \>
  \varphi_M\bigg({\bf r}-\frac{{\bf r}'}{2}\bigg).
\end{equation}
It is convenient to change coordinates again to 
\begin{align}
  {\bf R}^{(\prime)} =({\bf r}^{(\prime)}_1+
  {\bf r}^{(\prime)}_2)/2, \\
  {\bf r}^{(\prime)} ={\bf r}^{(\prime)}_1- {\bf r}^{(\prime)}_2
\end{align}
with conjugated momenta
\begin{align}
  \tilde {\bf P} &= {\bf p}_1 + {\bf p}_2, \\
  {\bf q} &= ({\bf p}_1-{\bf p}_2)/2.
\end{align}
  
We arrive at
\begin{multline}
  \label{eq:res1}
  N_M = \sum_{ab}\int \frac{d^3 P}{(2\pi)^3} \frac{d^3 q}{(2\pi)^3}
  \int d^3 R \, d^3 r \, d^3 r' \> \\
  \times W_{ab}\bigg( {\bf R}+\frac{\bf r}{2},
  {\bf R} - \frac{\bf r}{2}; \frac{\bf P}{2}+{\bf q},\frac{\bf P}{2}-{\bf q} 
  \bigg) \\ \times e^{i {\bf q}\cdot {\bf r}'} \varphi_M\bigg( {\bf r} + 
  \frac{{\bf r}'}{2}\bigg) \varphi_M^* 
  \bigg( {\bf r} -\frac{{\bf r}'}{2} \bigg).
\end{multline}
The integration over ${\bf R}'$ has been carried out and provides the 
three-momentum conservation ${\bf P} = \tilde{\bf P}$.

We define the Wigner function $\Phi^W_M$ of the meson as
\begin{equation}
  \Phi^W_M ({\bf r},{\bf q}) = \int d^3 r' e^{-i {\bf q}\cdot {\bf r'}} 
  \varphi_M\bigg( {\bf r} + \frac{{\bf r}'}{2}\bigg) 
  \varphi_M^* \bigg( {\bf r} -\frac{{\bf r}'}{2} \bigg) .
\end{equation}
Then 
\begin{multline}
 \frac{d N_M}{d^3 P} =  (2\pi)^{-3} \sum_{ab}\int d^3 R \int \frac{d^3 q \> 
  d^3 r}{(2\pi)^3} \\ \times
  W_{ab}\bigg( {\bf R}+\frac{\bf r}{2},{\bf R} - 
  \frac{\bf r}{2}; \frac{\bf P}{2}+{\bf q},\frac{\bf P}{2}-{\bf q} \bigg) 
  \> \Phi^W_M ({\bf r},{\bf q}).
\end{multline}

To evaluate this expression, we have to model the Wigner functions of
the parton system and of the meson. 
We assume that the 2-parton Wigner function can be factorized into a 
product of classical one-particle phase space distributions $w$. Furthermore
the color and helicity states for each flavor will be degenerate. We can 
therefore replace the sum over quantum numbers $a$ and $b$ by a degeneracy 
factor $C_M$.

We introduce the following simplifications:
the spatial width $\Delta {\bf r}$ of the hadron wave function, translating
into a width of $\Phi^W_M$, will be small compared to the nuclear size of 
the system at hadronization. The phase space distributions 
$w({\bf r};{\bf p})$ are steeply falling functions of the momentum 
${\bf p}$, but vary much less with the spatial coordinate ${\bf r}$ within 
the typical size $\Delta {\bf r}$ of a hadron. We shall assume that the 
spatial variation of the phase space distribution is small on this scale, 
replacing ${\bf R}\pm {\bf r}/2$ with ${\bf R}$.

In our derivation we implicitly chose an equal time formalism by
introducing the spatial coordinate states $|\hat {\bf r} \rangle$ at
a fixed time. However, in our result the integration over the
final state phase space 
\begin{equation}
  d^3 P d^3 R = d^3 P d^3 R \> P\cdot u(R)/ E
\end{equation}
is manifestly Lorentz invariant.
Here $E$ is the energy of the four vector $P$ and $u(R)$ is the future oriented
unit vector orthogonal to the hypersurface defined by the hadronization volume.
This form can be easily generalized for an arbitrary hadronization
hypersurface $\Sigma$ \cite{CoFr:74,ScheiHei:99}. We have
\begin{multline}
  E \frac{d N_M}{d^3 P} = {C_M} \int\limits_\Sigma 
  \frac{d^3 R \> P\cdot u(R)}{(2\pi)^3}  \int \frac{d^3 q}{(2\pi)^3} \>  
  \\ \times
  w_a\bigg( {R} ; \frac{\bf P}{2}-{\bf q} \bigg)
   \> \Phi^W_M ({\bf q}) \>
  w_b\bigg( {R}; \frac{\bf P}{2}+{\bf q}
  \bigg).
\end{multline}
$a$, $b$ now only denote the flavors of the valence quarks in meson $M$ and
\begin{equation}
  \Phi^W_M ({\bf q}) = \int d^3 r \> \Phi^W_M ({\bf r},{\bf q})
\end{equation}
is the spatially integrated Wigner function of the meson.

\subsection{Local light cone coordinates}

The structure of hadrons is best known in the infinite momentum frame which 
is described in light cone coordinates. If we let the hadron momentum 
${\bf P}$ define the $z$ axis of the hadron light cone (HLC) frame, we can 
introduce the light cone coordinates e.g. $q^+$, $q_\perp$ 
for the relative momentum $q$. Note that we denote transverse momenta
in the HLC frame --- i.e.\ the component orthogonal to ${\bf P}$ --- by the 
label $\perp$, but transverse momenta in the center of mass (CM) frame of the 
heavy ion collision  --- orthogonal to the beam axis --- by the label $T$. 
The labels $+$ and $-$ always refer to light cone coordinates in the HLC frame 
for a given ${\bf P}$. We fix the HLC frame by a simple rotation from the CM 
frame, i.e.\ $P_+ = (E+|{\bf P}|)/\sqrt{2}$.
We reintroduce the momentum  $k = P/2 - q$ of parton $a$ in the meson. 
Assuming a mass shell condition $k^2 = m_a^2$ with arbitrary but fixed 
virtuality $m_a$, we can rewrite the integral
\begin{equation}
  d^3 k = d^4 k \> 2 k^0 \delta(k^2-m_a^2) = d k^+ d^2 k_\perp
  \> \frac{k^0}{k^+}
\end{equation}
in HLC coordinates. $m_a$ will be of order $\Lambda_\text{QCD}$, 
or more precisely, of the order of a constituent quark mass.
In the HLC frame, where we assume that formally
$P^+ \to \infty$, we parametrize $k^+ = x P^+$ with $0\le x \le 1$.
Since $k^- \ll k^+$, we have $k^0 / k^+ \approx 1/\sqrt{2}$.
We end up with
\begin{widetext}
\begin{equation}
  E \frac{d N_M}{d^3 P} = {C_M} \int\limits_\Sigma 
  \frac{d^3 R \> P\cdot u(R)}{(2\pi)^3} 
  \int \frac{d x P^+ d^2 k_\perp}{\sqrt{2} (2\pi)^3} \>
  w_a\big( {R} ; x P^+ , {\bf k}_\perp \big)
   \> \Phi_M (x,{\bf k}_\perp) \>
  w_b\big( {R}; (1-x) P^+ , -{\bf k}_\perp \big).
\end{equation}
\end{widetext}
Here we have rewritten the spatially integrated Wigner function of the meson 
in terms of light cone coordinates for quark $a$ and approximated it by a
squared light cone wave function of the meson
\begin{equation}
  \Phi_M (x,{\bf k}_\perp) = \big| \bar\phi_M(x,{\bf k}_\perp) \big|^2.
\end{equation}

The light cone wave functions we introduce here do not necessarily 
coincide with the light cone wave functions 
$\Phi_{\rm LC} (x,{\bf k}_\perp,\mu)$ used for exclusive processes in QCD
\cite{ERBL:old}. 
There, a hadron is decomposed into a series of Fock states of perturbative 
partons, starting from the valence structure. In terms of this expansion 
we know that the valence Fock state has only a small contribution at scales 
$\mu>1\text{ Ge}V$.
$\mu$ is given by the momentum transfer in the hard reaction which is
described by perturbative QCD (pQCD). 
However, because the momentum transfer in a hard exclusive reaction has to 
be spread over all partons in the hadron state, the exclusive process itself 
acts like a filter, weighting the lower Fock states more strongly. 
In other words, 
the contributions of higher Fock states, though more likely in the wave 
function, are generally suppressed by inverse powers of the momentum transfer 
(higher twist). This usually permits a fairly good description of hard 
exclusive processes in terms of the lowest Fock state.

As we will discuss below, we expect the parton spectrum of a heavy ion 
collision at freeze out to be composed of an exponential part at small 
transverse momentum and a power law tail, given by pQCD, at high transverse
momentum. We will study recombination in the pQCD domain in a forthcoming
publication. 
%We expect it to be a small effect, as it is an exclusive 
%process, compared to the fragmentation. 

Here, we want to focus on recombination from the exponential part of the
parton spectrum. It will be given by a slope $1/T^*$ with a temperature-like 
parameter $T^*$. $T^*$ also sets the scale for the typical momentum transfer 
in the parton medium before hadronization. If $T^*$ is an effective blue 
shifted temperature $T^*=\sqrt{(1+\beta)/(1-\beta)}T$ in an expanding medium 
with physical temperature $T$ and flow velocity $\beta$, then the typical 
scale $T < T^*$ will even be smaller. Here we 
use the phase transition temperature at zero baryon density 
$T \approx 175 \text{ MeV}$ \cite{Karsch:01}.

We cannot expect that perturbative QCD will work as a description of partons
at the phase transition \cite{KaKaLaLu:99}.
What are the quanta that recombine? We know that in pQCD for 
decreasing scales the non-perturbative (long-range) matrix elements describing 
hadrons get more ``valence like'', though we cannot seriously extend this 
study to scales below 1 GeV. We will assume here that we recombine effective
constituent partons, taking into account only the valence structure
of the hadron. Gluons are no dynamic degrees of freedom in this picture,
and the quarks and antiquarks will have an effective mass. 

This picture is supported by the recent discovery of ``magical factors'' of 2 
and 3 in measurements of spectra and the elliptic flow of mesons and baryons 
respectively at RHIC \cite{Sorensen:03}. Later we will apply 
our assumptions to partons in the exponential spectrum having as much as 
2 GeV/$c$ of transverse momentum. This might raise doubts about the validity 
of an effective description. However, we have to keep in mind that the 
momentum is in principle meaningless, only the
momentum transfer experienced by a particle in a reaction 
sets the scale at which we resolve its structure. 
%So even with a 2 GeV/$c$ 
%quark the question comes back to what is the typical momentum transfer 
%in the parton phase at hadronization --it will be of order $T^*$. 

From the normalization condition 
\begin{equation}
  \langle M; {\bf P} | M; {\bf P}' \rangle = 
  (2\pi)^3 \delta({\bf P} - {\bf P}')
\end{equation}
we infer  $\int d^3 r \varphi_M(r) = 1$ and finally
\begin{equation}
  \label{eq:lcnorm}
  \int\frac{dk^+ d^2 k_\perp}{\sqrt{2} (2\pi)^3} \>
  |\bar\phi_M(x,{\bf k}_\perp)|^2 = 1.
\end{equation}
This is  different from the light cone wave functions in exclusive 
processes which involve a dimensionful quantity connected to the weight of
the Fock state, e.g.\ the pion decay constant $f_\pi$ in case of the pion.  

We further utilize the fact that we work in a kinematic regime where
$P^+$ is large compared to all non-perturbative quantities. Of course, our
main concern here is that we do not really know the shape of the wave 
function $\phi_M$. 
We can choose a factorized ansatz
\begin{equation}
  \bar\phi_M(x,k_\perp) = \phi_M(x) \Omega(k_\perp)  
\end{equation}
with a longitudinal distribution amplitude $\phi_M(x)$ and a transverse
part. We know that the transverse shape should be quite narrow, e.g.\ given
by a Gaussian with a width $\Lambda_\perp \ll P^+$.
Considering hadrons at mid rapidity, ${\bf k}_\perp$ will mainly be pointing 
in the longitudinal and azimuthal directions in the CM frame, where the 
variation of the parton distributions $w$ is small. 
This implies $w(R;x P^+, {\bf k}_\perp) \approx w(R;x P^+)$ for typical
transverse momenta ${\bf k}_\perp \sim \Lambda_{\text{QCD}}\ll P^+$.
We can then integrate the ${\bf k}_\perp$ dependence of the wave 
function. This leaves us with
\begin{widetext}
\begin{equation}
  \label{eq:meslc}
  E \frac{d N_M}{d^3 P} = {C_M} \int\limits_\Sigma 
  \frac{d^3 R \> P\cdot u(R)}{(2\pi)^3} 
  \int\limits_0^1  {d x} \>  
  w_a\big( {R} ; x P^+ \big)
   \> \left| \phi_M (x) \right|^2 \>
  w_b\big( {R}; (1-x) P^+ \big).
\end{equation}
\end{widetext}

The amplitude $\phi_M (x)$ encodes the remaining QCD dynamics. We expect 
it to be peaked around $x=1/2$, meaning that the two quarks will carry roughly
the same amount of momentum. But the width of the distribution, since it
is formulated in terms of momentum fractions, could be quite broad in momentum
space. Thus we cannot use the same argument as for the transverse
coordinates in order to integrate out this degree of freedom. However, for an
exponential parton spectrum we have
\begin{multline}
  w_a\big( {R} ; x P^+ \big) w_b\big( {R}; (1-x) P^+ \big) \\
  \sim e^{-xP^+/T} e^{-(1-x)P^+/T} = e^{-P^+/T}.
\end{multline}
Hence the product of parton distributions is independent of $x$ and we
can perform the integral over $x$, which just gives the trivial
normalization of the wave function from (\ref{eq:lcnorm}) \cite{FMNB:03}.
There will be corrections to that from momentum components other than $P^+$
which are not additive because energy is not conserved (see next subsection). 
Where we want to take into account wave functions, we adopt the asymptotic 
form of the perturbative pion distribution amplitude 
\begin{equation}
 \phi_M(x) = \sqrt{30} x(1-x)
\end{equation}
as a model.

\subsection{Energy conservation}

Energy conservation is not manifest in the recombination approach. 
Since we are dealing with a 2$\to$1 process, one of the particles in
general needs to be off mass shell \cite{ScheiHei:99}.
This does not pose a problem in the physical environment where recombination
takes place. Both in the quark phase and in the hadronic phase we expect 
interactions with the surrounding medium to occur. 
Since recombination, as described in this work, has been simplified to
a counting of quantum numbers and momenta, without real QCD
dynamics, we neglect effects of these additional interactions that ensure 
energy conservation.
This is tolerable due to the small time scale of hadronization
and that changes in the parton distributions $w$ by interactions during
this time are negligible. 
We also neglect effects that final state interactions
between the hadrons could have on the spectra -- these are expected 
to contribute mainly in the low momentum domain, $P_T < 1$~GeV.

For large momenta ${\bf P}$, the energies 
of the particles are dominated by the kinetic energies and not by the masses.
The light cone wave functions have a transverse width $\Lambda_\perp$
which is a non-perturbative momentum scale. Therefore the momenta of the
recombining quarks will be collinear up to transverse momenta of order
$\Lambda_\perp$. This implies that the energy is conserved up to terms
of order $\Lambda_\perp^2 / P^+ $ and $m^2/ P^+$, where $m$ stands
for the masses of the participating particles. 

Energy conservation is a problem at low transverse
momentum and can only be overcome if one takes  further 
interactions between the partons into account. 
This would require a much more sophisticated
formulation that includes nonperturbative initial- and final-state effects.

Restricting ourselves to large 
transverse momentum finally permits to use light
cone fractions $x$ for the spatial momentum ${\bf P}$ in the CM frame instead
of the momentum $P^+$ in the HLC frame. The relation
$p^+ = x P^+$ between the quark momentum $p$ and the hadron momentum $P$ in
light cone coordinates translates to 
\begin{equation}
  {\bf p} = x {\bf P} + \mathcal{O}\left( x\frac{M^2}{|{\bf P}|}\right)
  + \mathcal{O}\left( \Lambda_\perp \right).
\end{equation}
This is a leading order expansion in $M^2/{\bf P}^2$ where $M$ is the hadron 
momentum. At mid rapidity this further translates to the simple formula 
$p_T = x P_T$ for the transverse momenta.

\subsection{Low transverse momentum and hadron thermodynamics}

It is well known that total hadron yields at RHIC can be described 
very accurately by a purely statistical model, using only hadronic
properties, such as masses, chemical potentials and spin degeneracies 
\cite{BroFlo:01,PBmMaReSt,XuKa}. The hadron yields are given by the 
$P_T$-integrated spectra and are naturally dominated by particles with less 
than 2 GeV/$c$ transverse momentum. The picture of thermal hadron production 
does not necessarily require the existence of a parton phase.

However, one can show that recombination of a thermalized 
parton phase is consistent with thermal hadron production in the limit 
$P_T \to \infty$. Although pQCD will eventually dominate over both
mechanisms at large $P_T$, this nevertheless suggests that the recombination 
mechanism connects a thermal parton phase with the observed thermal hadron 
phase. Therefore it is justified to call recombination from a thermal parton
phase the microscopic manifestation of statistical hadron production.

\subsection{Summary of the formalism}

We want to summarize what we have so far. From  (\ref{eq:meslc})
the meson spectrum is given by 
\begin{widetext}
\begin{equation}
  \label{eq:res3}
  E \frac{N_M}{d^3 P} = C_M \int\limits_\Sigma d\sigma_R 
  \frac{P\cdot u(R)}{(2\pi)^3} \int\limits_0^1  {d x} \>  
  w_a\big( R ; x {\bf P} \big)
   \> \left| \phi_M (x) \right|^2 \>
  w_b\big( R; (1-x) {\bf P} \big).
\end{equation}
where $d\sigma_R$ measures the volume of the hypersurface $\Sigma$
and $R \in \Sigma$. For baryons, the same steps result in the expression
\begin{equation}
  \label{eq:protres}
  E \frac{N_B}{d^3 P} = C_B \int\limits_\Sigma d\sigma_R 
  \frac{P\cdot u(R)}{(2\pi)^3}
  \int \mathcal{D}x_i \, w_a\big( R ; x_1 {\bf P} \big)
  w_b\big( R; x_2 {\bf P} \big) 
  w_c\big( R; x_3 {\bf P} \big)
   \> \left| \phi_B (x_1,x_2,x_3) \right|^2 .
\end{equation}
\end{widetext}
$a$, $b$ and $c$ are the valence partons and $\phi_B(x_1,x_2,x_3)$ is the 
effective wave function of the baryon in light cone coordinates. 
We use the short notation
\begin{equation}
  \int \mathcal{D}x_i = \int\limits_0^1 {d x_1 \, d x_2 \, d x_3} 
  \delta(x_1+x_2+x_3-1)
\end{equation}
for the integration over three light cone fractions. Inspired 
by the asymptotic form of the light cone distribution amplitudes for 
pions and nucleons we choose
\begin{align}
  \label{eq:barlcwf}
  \phi_M(x) &= \sqrt{30} x(1-x), \\
  \phi_B(x_1,x_2,x_3) &= 12\sqrt{35} x_1 x_2 x_3.
\end{align}

These wave functions are broad in momentum space.
In order to study the effect of the width of the wave functions on our
results it will be interesting to alternatively 
explore the case of narrow wave functions in the spirit of Sec.\ 
\ref{sec:simple}. The limiting case are $\delta$-shaped wave functions
\begin{align}
  \left| \phi_M(x)\right|^2  &= \delta\left(x-\frac{1}{2}\right) \\
  \left| \phi_B(x_1,x_2,x_3)\right|^2 &= 
  \delta\left(x_1-\frac{1}{3}\right) \delta\left(x_2-\frac{1}{3}\right).
\end{align}
The spectra are then given by
\begin{equation}
  \label{eq:nwres}
  E \frac{N_M}{d^3 P} = C_M \int\limits_\Sigma d\sigma_R 
  \frac{P\cdot u(R)}{(2\pi)^3}  \>  
  w_a\Big( R ; \frac{\bf P}{2} \Big) 
  w_b\Big( R; \frac{\bf P}{2} \Big) 
\end{equation}
\begin{multline}
  \label{eq:nwres2}
  E \frac{N_B}{d^3 P} = C_B \int\limits_\Sigma d\sigma_R 
  \frac{P\cdot u(R)}{(2\pi)^3}  \>  \\ \times
  w_a\Big( R ; \frac{\bf P}{3} \Big) 
  w_b\Big( R; \frac{\bf P}{3} \Big)
  w_c\Big( R; \frac{\bf P}{3} \Big).
\end{multline}
It is an important observation that in the case of exponential parton 
distributions, the shape of the wave function is almost negligible.
We have to be aware that there will be corrections to the above equations
of order $m/|{\bf P}|$ and $\Lambda_{\rm QCD}/|{\bf P}|$, where $m$ is
the mass of the hadron or the partons, reducing their range of applicability
to large ${\bf P}$.

On the other hand the spectrum of hadrons from fragmentation is given by 
(\ref{eq:fracmaster})
\begin{equation}
  \label{eq:frac2}
  E \frac{d N_h}{d^3 P} = \sum_a \int\limits_0^1 \frac{d z}{z^2} 
  D_{a\to h}(z) E_a \frac{d N_a}{d^3 P_a}.
\end{equation}
The number of partons can be obtained from the cross section via the impact
parameter dependent nuclear thickness function 
$dN/d^3 P = T_{\text{AuAu}}(b) d\sigma_R/d^3 P$. We take $T_{\text{AuAu}}(0) = 
9A^2/8\pi R_A^2$ for central collisions. $R_A = A^{1/3} 1.2\text{ fm}$ is the
radius of the nucleus.

%The overall normalization is fixed by the particle number of the partons on
%the hypersurface $\Sigma$
%\begin{equation}
%  \label{eq:onepartnorm} 
%  N_a = \int \frac{d^3 p}{(2\pi)^3 p^0} d \sigma \> p\cdot u  w_a(R;p).
%\end{equation}

\section{Hadron and parton spectra}

In the following we want to discuss the parton phase created
in collisions of gold nuclei at RHIC with $\sqrt{s}=200$ GeV per nucleon pair.
We will then proceed to calculate 
the hadron spectra emerging from recombination and 
fragmentation of this parton phase.

\subsection{Modeling the parton phase}

Assuming longitudinal boost invariance, we fix the hypersurface $\Sigma$ by 
choosing $\tau = \sqrt{t^2 - z^2} = {\rm const.}$ for 
\begin{multline}
  R^\mu = (t,x,y,z) \\ =
  (\tau \cosh \eta, \rho \cos \phi,
  \rho \sin \phi, \tau \sinh \eta).
\end{multline}
It is convenient to introduce the space 
time rapidity $\eta$ and the radial coordinate $\rho$, since the measure
for the hypersurface $\Sigma$ then takes the simple form
$d\sigma_R = \tau \, d\eta \, \rho \, d\rho \, d\phi$
and the normal vector is given by
\begin{equation}
  u^\mu(R) = (\cosh \eta, 0, 0, \sinh\eta).
\end{equation}
Using a similar parametrization of the parton momentum 
\begin{equation}
  p^\mu = (m_T \cosh y, p_T \cos\Phi, p_T \sin\Phi, m_T \sinh y)
\end{equation}
with rapidity $y$ and transverse mass $m_T = \sqrt{m^2 + p_T^2}$, we
obtain $p\cdot u(R) = m_T \cosh(\eta-y)$. We remind the reader that we
will use capitalized variables ($P, P_T, M, M_T$, etc.) for hadrons.

Now we have to specify the spectrum of partons. As already discussed above we 
assume that the parton spectrum consists of two domains.
At large $p_T$, the distribution of partons is given by perturbative QCD and 
follows a power law. For the transverse momentum distribution at midrapidity 
for central collisions (b=0) we use the parametrization \cite{FMS:03,SGF:03}
\begin{equation}
  \frac{d N^{\text{pert}}_a}{d^2 p_T dy}\Big|_{y=0} = 
  K \frac{C}{(1+p_T/B)^\beta}.
  \label{eq:pqcdpart}
\end{equation}
The parameters $C$, $B$, and $\beta$ are taken from a leading order (LO) pQCD
calculation and can be found in \cite{SGF:03} for the three light quark
flavors and gluons. A constant $K$ factor of 1.5 is included to roughly 
account for higher order corrections in $\alpha_s$ \cite{EskTuo:01}.
The calculation includes nuclear shadowing of the parton distributions, 
but no higher twist initial state effects. Higher twist effects, like the 
Cronin effect, will fade like $A^{1/3} \Lambda^2_{\text{QCD}}/P_T^2$ for 
high transverse momentum \cite{LQS:92}. Since we will show that 
fragmentation and pQCD are only dominant for transverse momenta above 
5 GeV/$c$ in the hadron spectrum for RHIC, it is safe to omit the Cronin 
effect.

\begin{figure}
%  \begin{center}
  \epsfig{file=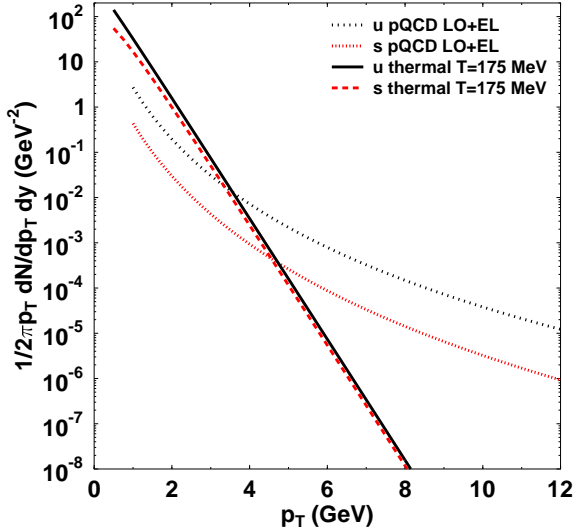,width=0.9\columnwidth}
%  \end{center}
  \caption{\label{fig:partonspec} Spectrum of $u$ and $s$ quarks at 
     hadronization in a central Au+Au collision at RHIC.
     Perturbative partons from hard QCD processes with subsequent energy loss 
     (dashed lines) and the thermal phase with $T=175$ MeV and radial flow
     $v_T = 0.55 c$ (solid lines) are shown.}
\end{figure}

Energy loss of partons, resulting
in a shift of the transverse momentum spectrum \cite{BDMS:01,Muller:02},
is taken into account and parameterized as
\begin{equation}
  \Delta p_T(b,p_T) = \epsilon(b) \, \sqrt{p_T} \, 
    \frac{\langle L \rangle}{R_A} .
  \label{eq:enloss}
\end{equation}
For central collisions we take $\langle L\rangle = R_A$ and therefore
$ \Delta p_T(0,p_T) = \epsilon(0) \sqrt{p_T}$ (we postpone the discussion of
the impact parameter dependence to Subsection IV.C).   The choice of 
$\langle L\rangle = R_A$ neglects the fact that for the strong quenching 
observed at RHIC energies, jet emission becomes a surface effect 
\cite{Muller:02,Shuryak:02}. However, we note that only the product 
$\epsilon\langle L\rangle$ as a whole is a parameter. We have no ambition here
to make a connection to the microscopic parameters of jet quenching, therefore
we will not disentangle $\epsilon$ and $\langle L\rangle$.
This would require a more sophisticated model of the emission geometry.
We also do not use a radial profile for the emission and the density of 
the medium. This can be found discussed elsewhere in the literature 
\cite{Muller:02,WaWa:02,Wang:03}.
Nevertheless, our ``minimal'' description of energy loss is quite successful
to describe the available data on high-$P_T$ spectra ($P_T > 5$ GeV/$c$) for 
$\pi^0$, $K^0_s$ and charged hadrons $(h^+ + h^-)/2$ in central Au+Au 
collisions. From a fit to these data we find 
$\epsilon_0=\epsilon(0)=0.82\text{ GeV}^{1/2}$ for
central Au+Au collisions. This value corresponds to an average energy loss 
of 3 GeV for a 10 GeV parton in a central Au+Au collision. 
The perturbative spectrum for
up and strange quarks at midrapidity is shown in Fig.\ \ref{fig:partonspec}.
For fragmentation of pions, kaons, protons and antiprotons we use LO KKP 
fragmentation functions with the scale set to the hadron transverse momentum 
$P_T$ \cite{KKP:00}. $\Lambda$ fragmentation is calculated with the LO 
fragmentation functions of de Florian, Stratmann and Vogelsang \cite{dFSV:97}.

%Compared to more sphisticated studies of jet quenching \cite{Muller:02,
%and more} we do not keep the full geometry which would require to have
%the point of emission of the jet and the azimuthal angle of the emission 
%%as degrees of freedom that have to be integrated over. We also do not use 
%a radial profile for the emission and the density of the medium. 
%Instead $\langle L\rangle$ is taken to be an effective averaged length that 
%the parton has to travel through the hot medium and $\epsilon$ determines the 
%amount of the energy loss. However, both $\langle L\rangle$ and $\epsilon$ 
%depend on the impact parameter $b$. 
%This approach does not use the fact the 
%hadron emission becomes a surface effect in central collisions at RHIC 
%energies, but it is sufficient for our purposes here. The description of the 
%data is surprisingly good. In order to connect the microscopic 
%parameters of jet quenching to the observed energy loss, however, it is
%necessary to keep the more sophisticated space time picture. This
%is done elsewhere in the literature \cite{Muller:02,and others}.

Besides the perturbative tail of the parton spectrum that will turn into 
hadrons via fragmentation, we assume the existence of a spectrum of 
thermalized partons that are recombining at hadronization and dominate 
at low and intermediate values of $p_T$. 
In this phase we assume the effective degrees of freedom to be 
constituent quarks without dynamical gluons. We take the spectrum to be
exponential with a given temperature $T$
\begin{equation}
  \label{eq:partondistr1}
  w_a(R;p) = \gamma_a e^{-p\cdot v(R)/T} e^{-\eta^2/2\Delta^2}
  f(\rho,\phi).
\end{equation}
$\gamma_a$ is a fugacity factor for each parton species $a$. We also include 
longitudinal and radial flow through the velocity vector
\begin{multline}
  \label{eq:flowvel}
  v^\mu (R) = (\cosh \eta_L \cosh\eta_T, \sinh\eta_T \cos\phi, \\
  \sinh\eta_T \sin\phi, \sinh\eta_L\cosh\eta_T).
\end{multline}
$\eta_L(R)$ and $\eta_T(R)$ are the rapidities of the longitudinal
and radial flow which still could depend on the space time point $R\in\Sigma$.
For the longitudinal expansion we choose a Bjorken scenario where the 
longitudinal rapidity is simply fixed by the space time rapidity
\begin{equation}
  \label{eq:longflowrap}
  \eta_L(R) = \eta.
\end{equation}
The transverse flow is given by a velocity $v_T(R)$ with
$v_T = \tanh \eta_T$. For practical purposes we will not work with a radial
profile but assume $v_T$ to be independent of $\rho$ and $\phi$. However, for 
collisions with finite impact parameter $b$ we will later allow a dependence 
of $v_T$ on the azimuthal angle $\phi$ in order to describe the measured 
elliptic asymmetry in the spectra. The space-time structure of the parton 
source in (\ref{eq:partondistr1}) is given by a transverse distribution 
$f(\rho,\phi)$ and a wide Gaussian rapidity distribution with a width 
$\Delta$. 

We assume that hadronization occurs at $\tau = 5$ fm at a temperature
$T=175$ MeV in the parton phase. This is consistent with predictions of the 
phase transition temperature at vanishing baryon chemical potential from
lattice QCD \cite{Karsch:01}. 
For the spread of the parton distribution in longitudinal direction we choose 
$\Delta=2$. The constituent quark masses are taken to
be 260 MeV for $u$ and $d$ quarks and 460 MeV for $s$ quarks.

The two component model of the parton spectrum with an exponential bulk and
a power law tail is also predicted by parton
cascades like {\sc VNI/BMS} \cite{BMS:02}, although the interactions in that
case are purely perturbative. This implies that an exponential shape of the 
spectrum does not necessarily mean that the parton system is in thermal 
equilibrium.

In the region where contributions from recombination and fragmentation are
of the same size we expect other mechanisms to play a role, which interpolate
between the two pictures. This could include partial recombination and 
higher twist fragmentation.  In the absence of a consistent description 
of these mechanisms we simply add both contributions to the hadron spectrum 
-- recombination from the exponential part and the fragmentation from the 
pQCD part -- for $P_T > 2$ GeV/$c$.

\subsection{Degeneracy factors}

It is not a priori clear from QCD what the degeneracy factors $C_h$ for
each hadron $h$ are. In principle every quark has 3 color and 2 spin degrees
or freedom. On could argue that 3 quarks of any color and spin can form
a proton and that quantum numbers can be "corrected" at no cost by the 
emission of soft gluons. That would lead to degeneracy factors
$C_p = 2\times(3\times2)^3$ and $C_{\pi^0} = (3\times2)^2$. On the other hand 
there are no dynamical gluons in our picture and it would be consistent to 
require recombining partons to have the right quantum numbers at the beginning.

Surprisingly this is supported from work on recombination in pQCD, where
the contributions from color octets and spin-flip states to the recombination
of $D$ mesons were found to be small \cite{BraJiaMe:02}. Using this assumption,
the degeneracies are only determined by the degrees of freedom of the
hadron, e.g.\ $C_p=2$, $C_{\pi}=1$ etc. These are exactly the degeneracies
used in the statistical thermal model. We will not take into account 
feeddown from decays of resonances, except for the $\Lambda$, where the
$\Sigma^0$ is too close in mass to be suppressed. Hence we use $C_\Lambda = 4$.
 
This is different from \cite{FMNB:03} where we counted $\Delta$ resonances
to give nucleons with weight 1. However this overestimates the 
correction from $\Delta$ decays. Nevertheless the degeneracy of $5/3$ given 
in \cite{FMNB:03} for the proton was of the right size due to a
mistake in the normalization of the baryon states, that gave an additional 
factor of $1/3!$. Therefore the numerical results given in \cite{FMNB:03} are 
still valid.

We should add that due to the small but probably non-vanishing color octet
and spin flip contributions and due to feeddown corrections we expect all 
degeneracy factors to have an error of at least 20\%.

\subsection{Central collisions}

For the momentum spectrum of quarks
\begin{equation}
  E \frac{dN^{\text{th}}_a}{d^3 p} = g \gamma_a \int\limits_\Sigma d\sigma_R 
  \frac{p\cdot u(R)}{(2\pi)^3} w_a(R;p)
  \label{eq:partspec2}
\end{equation}
we rewrite the exponent in (\ref{eq:partondistr1}) as
\begin{multline}
  \label{eq:pdotu}
  p\cdot v(R) = 
  m_T \cosh(\eta-y) \cosh\eta_T \\
   - p_T \cos(\phi-\Phi)
  \sinh\eta_T 
\end{multline}
where $m_T$ is the transverse mass of quark $a$. The factor $g=6$
in  (\ref{eq:partspec2}) is counting the color and spin degeneracies.
In the case of central collisions (impact parameter $b=0$) we can 
assume that $v_T$ and $f$ are independent of $\phi$. For simplicity
we furthermore assume that $f$ is also independent of the radial coordinate
$\rho$ and that the radial distribution of the partons is homogeneous up
to a radius $\rho_0$: $f(\rho) = \Theta(\rho_0-\rho)$. 
We can then easily perform the $\phi$ and $\rho$ integrals in 
 (\ref{eq:partspec2}).
If we consider the region around $y=0$, we can also neglect the Gaussian 
profile function in $\eta$, since $\eta^2/\Delta^2 \ll (m_T/T)\cosh\eta$, 
and integrate over this variable analytically.

The transverse momentum spectrum of partons in the thermal phase is then given
by
\begin{multline}
  \frac{N^{\text{th}}_a}{d^2 p_T dy}\Big|_{y=0} = 
  2g\gamma_a m_T \frac{\tau A_T}{(2\pi)^3}  \\ \times  
  I_0 \left[ \frac{p_T \sinh \eta_T}{T}\right]
  K_1 \left[ \frac{m_T \cosh \eta_T}{T}\right].
  \label{eq:finalparton}
\end{multline}
$A_T=\rho_0^2 \pi$ is the transverse area of the parton 
system and $I_0$ and $K_1$ are modified Bessel functions.
Fig.\ \ref{fig:partonspec} also shows the thermal spectrum of up and strange
quarks. The radial flow velocity $v_T=0.55c$, fugacities $\gamma_u=\gamma_d=1$,
$\gamma_{\bar u}=\gamma_{\bar d}=0.9$, $\gamma_s=\gamma_{\bar s}=0.8$ and the 
radius $\rho_0=9$ fm were determined by fits of our calculation to the 
measured hadron spectra in central collisions, see below.

\begin{figure*}
  \epsfig{file=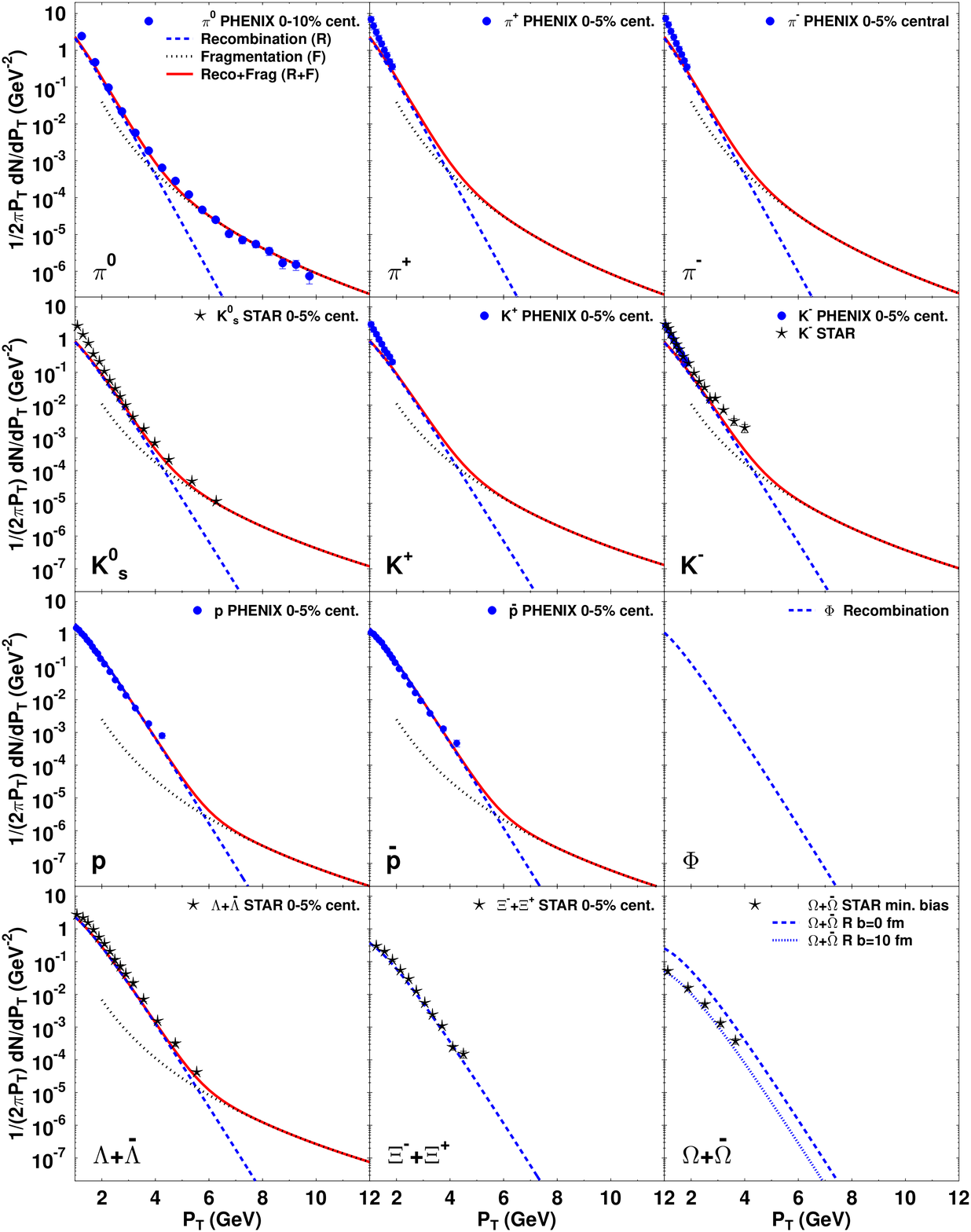,width=0.9\textwidth}
  \caption{\label{fig:spectra} Hadron spectra at midrapidity as a function of
    transverse momentum $P_T$ for central Au+Au collisions at $\sqrt{S}=200$ 
    GeV. We show fragmentation (dotted), recombination (dashed) and the sum
    of both contributions (solid line) versus data from PHENIX and STAR
    (for $\Phi$, $\Xi^+ + \Xi^-$ and $\Omega+\bar\Omega$ recombination only).
    $\Omega+\bar\Omega$ data is minimum bias, therefore the result
    of a calculation with impact parameter $b=10$ fm is also shown. All
    data is preliminary except $\pi^0$, $p$ and $\bar p$ data from PHENIX. 
    All error bars give
    statistical errors only, except for $\pi^0$ data from PHENIX which give the
    total error. See Sec.\ \ref{sec:results} for more details.}     
\end{figure*}

The transverse momentum spectrum of hadrons formed by recombination from
the thermal parton spectrum can be derived from
Eqs.\ (\ref{eq:res3},\ref{eq:protres}) to be
\begin{widetext}
\begin{equation}
  \frac{N_M}{d^2 P_T dy}\Big|_{y=0} = 
  C_M M_T \frac{\tau A_T}{(2\pi)^3} \, 2\gamma_a \gamma_b \,
  I_0 \left[ \frac{P_T \sinh \eta_T}{T}\right] 
  \int\limits_0^1 dx \> \left|\phi_M(x)\right|^2  k_M(x,P_T) \, ,
  \label{eq:messpec}
\end{equation}
\begin{equation}
  \frac{N_B}{d^2 P_T dy}\Big|_{y=0} = 
  C_B M_T \frac{\tau A_T}{(2\pi)^3} \, 2\gamma_a \gamma_b \gamma_c \,
  I_0 \left[ \frac{P_T \sinh \eta_T}{T}\right]
  \int \mathcal{D}x_i \,
  \left|\phi_B(x_1,x_2,x_3)\right|^2  
  k_B(x_i,P_T)
  \label{eq:barspec}
\end{equation}
for mesons and baryons respectively. We introduced the short notations
\begin{align}
  k_M(x,P_T) &=  
  K_1 \left[ \frac{\cosh \eta_T}{T}\left(\sqrt{m_a^2 + x^2 P_T^2}+
  \sqrt{m_b^2 + (1-x)^2 P_T^2}\right) \right]  \, , \\
  k_B(x_i,P_T) &=    
  K_1 \left[ \frac{\cosh \eta_T}{T}\left(\sqrt{m_a^2 + x_1^2 P_T^2}+
  \sqrt{m_b^2 + x_2^2 P_T^2} + \sqrt{m_c^2 + x_3^2 P_T^2} \right) \right].
\end{align}
\end{widetext}
Here $m_a$, $m_b$ and $m_c$ are the 
masses of the valence quarks and $M_T$ is the transverse mass of 
the hadron. Fig.\ \ref{fig:spectra} shows the result of our calculation 
for the spectra of pions, protons, antiprotons, kaons, Lambdas, Xis and Omegas.
See Sec.\ \ref{sec:results} for discussion.

%\twocolumngrid

\subsection{Peripheral Collisions}

In order to describe peripheral collisions, we have to scale the 
perturbative part of the parton spectrum given in (\ref{eq:pqcdpart}) 
by the ratio of thickness functions,
or equivalently by the number of binary nucleon collisions
\begin{equation}
  dN_a^{\text{pert}}(b)= \frac{T_{\text{AuAu}}(b)}{T_{\text{AuAu}}(0)} 
     dN_a^{\text{pert}} = \frac{N_{\text{coll}}(b)}{N_{\text{coll}}(0)} 
     dN_a^{\text{pert}}.
\end{equation}
Our values for the number of collisions $N_{\text{coll}}$ as a 
function of impact parameter are listed in Tab.\ \ref{tab:ncoll} and are
close to the values used by the PHENIX collaboration 
\cite{PHENIX:Adler03pi0auau}.

\begin{table}[h]
  \begin{ruledtabular}
  \begin{tabular}{c||c|c|c|c|c|c|c|c|c}
  $b$ [fm] & 0 & 5.5 & 7.5 & 9 & 10 & 11 & 12 & 13 &13.9 \\ \hline
  $N_{\text{coll}}$ & 1146 & 594 & 350 & 199 & 120 & 61.6 & 26.0 & 10.0 & 5.3
  \end{tabular}
  \end{ruledtabular}
  \caption{\label{tab:ncoll} Average number of binary nucleon  
  collisions $N_{\text{coll}}$ for some values of the impact parameter $b$ 
  in collisions of gold nuclei.}
\end{table}

The length and width of the overlap zone of two nuclei with radius $R_A$, 
colliding at impact parameter $b$, are
$l(b)=\sqrt{R_A^2 - (b/2)^2}$ and $w(b)=R_A-b/2$. We scale the average length
entering the energy loss in  (\ref{eq:enloss}) as $\langle L \rangle=
\left(l(b)+w(b)\right)/2$. On the other hand the density of the hot medium 
is decreasing with increasing impact parameter. We choose the simple ansatz
\begin{equation}
  \epsilon(b) = \epsilon_0 \frac{1-e^{-(2R_A -b)/R_A}}{1-e^{-2}}
  \label{eq:enlossbdep}
\end{equation}
for the $b$ dependence of the energy loss parameter, which describes the
data surprisingly well. We refer to \cite{Muller:02} for more detailed 
studies of the jet quenching effect.
%The $b$ dependence of the average energy loss for a 10 GeV parton is
%shown in Fig.\ \ref{fig:enlossbdep}.

For the thermal phase of the parton spectrum, we keep the temperature $T$
and the hadronization time $\tau$ independent of the impact parameter $b$,
but adjust the size of the volume according to the profile function 
$f(\rho)$. We scale the transverse area $A_T$ of the parton phase at 
hadronization with the transverse area of the overlap zone of the two nuclei
\begin{equation}
  A_T (b) = \frac{l(b)w(b)\pi}{R_A^2\pi}  A_T(0) = l(b)w(b) 
  \frac{\rho_0^2}{R_A^2}. 
\end{equation}
In principle, the radial flow velocity $v_T$ is expected to vary
with impact parameter. However, it turns out that the slope of the measured 
hadron spectra above 2 GeV/$c$ is consistent with a constant flow velocity 
$v_T=0.55 c$ up to very large impact parameters, so that we fix this value
for all $b$. It may be questionable whether partons are produced in
equilibrium in peripheral collisions. We therefore introduce an additional
impact parameter dependent fugacity $\gamma(b)$ common to all quark flavors.
However, the measured hadron spectra favor $\gamma(b)=1$ up to high values of
$b$. Corrections are only necessary for very peripheral collisions.
We take $\gamma(b)=1$ for $b\le 10.5$ fm, $\gamma(11\text{ fm})=0.7$,
$\gamma(12\text{ fm})=0.4$ and $\gamma(13\text{ fm})=0.4$.

\subsection{Elliptic Flow}

For hadrons from fragmentation we assume that the azimuthal anisotropy 
in peripheral collisions is induced by the azimuthal dependence of the 
energy loss \cite{Wang:00v2,GyViWa:00,Shuryak:02,Muller:02}. To determine 
the coefficient $v_2$ we generalize (\ref{eq:enloss}) to
\begin{equation}
  \Delta p_T(b,p_T,\Phi) = \epsilon_0 \sqrt{p_T} \frac{\langle L\rangle}{R_A}
  (1-\alpha \cos 2 \Phi), 
\end{equation}
where $\epsilon(b)$ is given by (\ref{eq:enlossbdep}). This leads to 
a spectrum $d^2N/dP_T^2$ with non trivial dependence on $\Phi$.
From that we obtain the elliptic flow by applying the definition
\begin{equation}
  v_2( P_T) = \langle \cos 2\Phi \rangle = \frac{\int d\Phi \, \cos 2\Phi \,
  d^2N/dP_T^2}{\int d\Phi \, d^2N/dP_T^2}.
\end{equation}
The parameter $\alpha$ is given by the collision geometry. It has the value
\begin{equation}
  \alpha = \frac{w(b)-l(b)}{w(b)+l(b)}
\end{equation}
for given impact parameter $b$.

In the thermal phase, the hydrodynamic expansion from an originally 
anisotropic overlap zone of both nuclei (for $b\ne 0$) induces an elliptic 
anisotropy of the parton spectrum. This is leading to a dependence of the 
transverse flow on the azimuthal angle $\phi$. To model the anisotropy in the 
parton phase we take the azimuthal dependence of the transverse rapidity to be
\begin{equation}
  \eta_T(\phi) = \eta_T^0 \left( 1- f(p_T) \cos 2 \phi \right) .
\end{equation}
Here $\eta_T^0$ is the rapidity given by the flow velocity
$v_T=0.55 c$, so that $\tanh \eta_T^0 = 0.55$. The amplitude of the
anisotropy $f(p_T)$ is given by the geometrical anisotropy $\alpha$ at low
transverse momenta, but faster partons will experience the anisotropy in the
expansion less than slower ones. Therefore we choose an ansatz
\begin{equation}
  f(p_T)= \frac{\alpha}{1+(p_T/p_0)^2}.
\end{equation}

From Eqs.\ (\ref{eq:partspec2},\ref{eq:pdotu}) one obtains
\cite{Schnedermann,HuKoHeRuVo}
\begin{widetext}
\begin{equation}
  v^a_2 (p_T)  =  \langle \cos (2 \Phi) \rangle 
%  = \frac{\int ^{2\pi}_{0} d \phi_p \int^{2\pi}_0  d \phi_s \cos (2 \phi_p) 
%  {\rm{e}} ^{  \frac{p_T \sinh \rho (\phi_s)}{T} \cos (\phi_s - \phi_p) }
%  K_1 \left(  \frac{m_T \cosh \rho (\phi_s)}{T}\right)} 
%  {\int ^{2\pi}_{0} d \phi_p \int^{2\pi}_0  d \phi_s 
%  {\rm{e}} ^{  \frac{p_T \sinh \rho (\phi_s)}{T} \cos (\phi_s - \phi_p) }
%  K_1 \left(  \frac{m_T \cosh \rho (\phi_s)}{T}\right)} \nonumber \\
  =  \frac{\int d \phi \, \cos 2\phi \>  I_2 \left[
  {p_T }\sinh\eta_T(\phi)/{T} \right] 
  K_1 \left[  {m_T }\cosh \eta_T (\phi)/{T} \right]}
  {\int d \phi \, I_0\left[
  {p_T} \sinh\eta_T(\phi)/{T} \right] 
  K_1 \left[  {m_T } \cosh \eta_T (\phi)/{T} \right]}.
  \label{eq:v2hydro} 
\end{equation} 
\end{widetext}
Our assumptions imply $v_2^u=v_2^{\bar u}=v_2^d=v_2^{\bar d}$ in the 
thermal phase, but $v_2^s=v_2^{\bar s}$ differ slightly at small 
$p_T$ because of the bigger strange quark mass. We determine the parameter
$p_0$ in the parametrization of the parton phase from a comparison to the 
PHENIX measurement of the elliptic flow of pions \cite{PHENIX:v2}. We obtain
$p_0 = 1.1$ GeV/$c$.

%This seems to be seen by the STAR collaboration \cite{Sorensen:03}. 

After having fixed the coefficient $v_2^a(p_T)$ for each parton species $a$ we
write the azimuthally anisotropic phase space distribution for thermal 
partons at midrapidity as
\begin{equation}
  w^{\text{ai}}_a(R;p) = w_a(R;p) \left[ 1+2v^a_{2}(p_T) \cos 2 \Phi \right].
  \label{eq:partonv2}
\end{equation}
Here $w_a(R;p)$ is the phase space 
distribution without anisotropy from Eq.\ (\ref{eq:partondistr1}).
Substituting this into the basic recombination formulae (\ref{eq:res3},
\ref{eq:protres}) we obtain
\begin{widetext}
\begin{equation}
  v^M_2 (P_T) = \frac{\int dx \,|\phi_M(x)|^2 \left[v^a_{2}\left(xP_T\right)
  +v^b_{2} \left((1-x)P_T\right) \right] k_M (x,P_T) }{
  \int dx \, |\phi_M(x)|^2 \left[ 1+ 2v^a_{2}\left(xP_T\right) v^b_{2}
  \left((1-x)P_T\right) \right]
  k_M (x,P_T)  }
  \label{eq:v2_1}
\end{equation}
\begin{equation}
  v^B_2 (P_T) = \frac{\int \mathcal{D}x_i \,
  |\phi_B(x_i)|^2 \left[ v^a_{2}
  \left(x_1 P_T\right)+v^b_{2} \left(x_2 P_T\right) +v^c_{2} 
  \left(x_3 P_T\right) + 3  v^a_{2}
  \left(x_1 P_T\right) v^b_{2} \left(x_2 P_T\right)  v^c_{2} 
  \left(x_3 P_T\right)\right] k_B(x_i,P_T) }{
  \int \mathcal{D}x_i \, |\phi_B(x_i)|^2 \left[ 1+ 2\left(
  v^a_{2}\left(x_1P_T\right) v^b_{2}
  \left(x_2P_T\right) + v^a_{2}\left(x_1P_T\right) v^c_{2}
  \left(x_3P_T\right) + v^b_{2}\left(x_2P_T\right) v^c_{2}
  \left(x_3P_T\right) \right) \right] k_B(x_i,P_T)  }  
  \label{eq:v2_2}
\end{equation}
\end{widetext}
for the anisotropies in the meson and baryon spectra respectively.
Using the $\delta$-function approximation this reduces to the relations
already given before in the literature \cite{MoVo:03,LinMol:03}
\begin{widetext}
\begin{align}
  v^M_{2} ( P_T )  =&  \frac{ v^a_{2}(\frac{1}{2}P_T)  + 
  v^b_{2}(\frac{1}{2}P_T)}
  {1+2v^a_{2}(\frac{1}{2}P_T)v^b_{2}(\frac{1}{2}P_T)},  \\
  v^B_{2} ( P_T )  =&  \frac{ v^a_{2}(\frac{1}{3}P_T) + 
  v^b_{2}(\frac{1}{3}P_T) + v^c_{2}(\frac{1}{3}P_T) + 
  3 v^a_{2}(\frac{1}{3}P_T) v^b_{2}(\frac{1}{3}P_T) v^c_{2}(\frac{1}{3}P_T)}
  {1+2v^a_{2}(\frac{1}{3}P_T)v^b_{2}(\frac{1}{3}P_T) 
  + 2v^b_{2}(\frac{1}{3}P_T)v^c_{2}(\frac{1}{3}P_T)
  + 2v^c_{2}(\frac{1}{3}P_T)v^a_{2}(\frac{1}{3}P_T)}.
\label{eq:v2bm}
\end{align}
\end{widetext}

If we assume one universal partonic $v_2$ for the recombining
quarks the above expressions simplify to
\begin{align}
  v_{2,M}(P_T)=&\frac{2 v_2(\frac{1}{2} P_T)}{1 + 2 v_2(\frac{1}{2} P_T)^2} 
  \>, \\
  v_{2,B}(P_T)=&\frac{3 v_2(\frac{1}{3} P_T) + 3 v_2(\frac{1}{3} P_T)^3 }
             {1 + 6 v_2(\frac{1}{3} P_T)^2} \>.
\end{align}
Since the maximal values of $v_2$ will be of the order of 0.1, we can
neglect the quadratic and cubic terms and arrive at the following
simple scaling law, which connects the elliptic flow of hadrons $v_2^h$
to those of the partons $v_2$:
\begin{equation}
  v^h_{2}(P_T)= n \, v_2\left(\frac{1}{n} P_T\right)
  \label{eq:v2scaling}
\end{equation}
with $n$ being the number of valence quarks and anti-quarks contained
in hadron $h$. This scaling law was indeed already found to hold in STAR data 
on the elliptic flow of $\Lambda$ and $K_s^0$ down to transverse momenta of 
about 500 MeV/$c$ \cite{Sorensen:03}. This is a very strong support for the 
recombination picture. Apparently a part of the uncertainty in the 
recombination mechanism at low $P_T$, introduced by the violation of energy 
conservation, cancels after taking the ratios in 
Eqs.\ (\ref{eq:v2_1},\ref{eq:v2_2}). The recombination formalism seems to give
valid results for $v_2$ down to transverse momenta of several hundred MeV/$c$.

%We include the azimuthal anisotropy in $\rho(\phi_s)$,  
%\begin{equation}
%\rho(\phi_s) = \rho_0 ( 1+ \alpha_p (p_T) \cos(2\phi_s)),
%\end{equation}
%where $\rho_0$ is described by the transverse flow $v_T$, 
%$\rho_0 = \frac{1}{2} \ln \frac{1+v_T}{1-v_T}$. 

%The anisotropy of flow, $\alpha(p_T)$ is determined geometrically and  
%has $p_T$ dependence, because the contribution of recombination 
%part to $v_2$ decreases with $p_T$.
%Hence we parametrize 
%\begin{equation}
%\alpha_p (p_T) = -\alpha_0 f(p_T), 
%\end{equation}
%where $f(p_T)$ is weight function, $f(p_T)= 1/(1+(p_T/p_0)^2)$, 
%and for fragmentation part it should be $1-f(p_T)$. 
%Using  (\ref{eq-v2parton}), we fix the parameter $p_0=1$ GeV for $v_2$ 
%of parton in order to reproduce the experimental data 
%\footnote{In order to apply the 
%experimental data to parton, we divide $v_2$ and $P_T$ by $n(=2,3)$.} and   
%then we calculate $v_2$ of baryon and meson, using  (\ref{eq-v2bm}).
%As before, we set the temperature $T=175$ MeV and the transverse flow 
%$v_T=0.55$.

We combine the contributions to the anisotropic flow from recombination and 
fragmentation by using the relative weight $r (P_T)$ for the recombination 
process
\begin{equation}
  v_2 (P_T) = r (P_T) v_{2,{\rm R}}(P_T) + \left(1-r(P_T)\right) 
  v_{2,{\text{F}}}(P_T).
\end{equation}
$r(P_T)$ is defined as the ratio of the recombination contribution to the
spectrum and the total yield.
\begin{equation}
  r(P_T) = \frac{dN^{\text{R}}/d^2 P_T}{(dN^{\text{R}}/d^2 P_T + 
  dN^{\text{F}}/d^2 P_T)}.
  \label{eq:recoweight}
\end{equation}

%The location of crossover of recombination and fragmentation in 
%$P_T$ affects $r(P_T)$, and it has centrality dependence.   

%From experimental data of PHENIX collaboration, the behavior of 
%elliptic flow as a function of centrality has a plateau 
%from 30 \% to 55 \% centrality \cite{Esumi}.  Over this centrality 
%range the deviation among the location of the crossover from 
%recombination to fragmentation in $P_T$ is small. Therefore we use 
%the ratio of recombination part in central collisions in analyses 
%of the elliptic flow.

\subsection{The statistical thermal model}

%We will also compare with the predictions of a purely hadronic 
%statistical model. In particular, we present results for hadron ratios, 
%and we will compare them with those from our calculations for 
%recombination and fragmentation. 
In this subsection we give a brief 
account of the statistical model following variant I of \cite{BroFlo:01}. 
For further details we refer the reader to the comprehensive literature 
\cite{BroFlo:01,PBmMaReSt,XuKa,zschiesche}.

The hadron spectrum at is supposed to emerge from a hypersurface $\Pi$
and has the form  
\begin{equation}
  E \frac{dN_h}{d^3P} = \int\limits_\Pi d \sigma_R
  \frac{ P\cdot v(R)}{(2\pi)^3} \, G_h(R;P).
\label{eq:particledist}
\end{equation}
We use the same parametrization for the four velocity $v(R)$ as in  
(\ref{eq:flowvel}). The hypersurface $\Pi$ is determined by the condition
$\sqrt{v^2} = \tau_{\text{SM}} = \text{const}$.
The hadronic phase-space distribution functions are given by  
\begin{equation}
  G_h(R;P) = \frac{C_h f_{\text{SM}}(r)}{e^{ -(P \cdot v - 
  \mu_B B_h - \mu_s S_h - \mu_I I_h)/T_{\text{SM}}} \pm 1} ,
\end{equation}
for bosons and fermions respectively. $r=\tau_{\text{SM}} \sinh \eta_T$ is
the radial coordinate and $f_{\text{SM}}(r)=\Theta(r_0-r)$ 
is a radial profile function providing a cylindrical shape.
$C_h$ is the degeneracy factor and $B_h$, $S_h$ and $I_h$ are 
baryon number, strangeness and third component of the isospin for hadron
species $h$.  

%where $\eta$ and $\phi_s$ are the rapidity and azimuthal angle in 
%coordinate space, respectively.
%The transverse size of system is described by $\alpha_\bot$ ($\rho = 
%\tau \sinh \alpha_\bot$) and the 
%transverse velocity ($v_T$) is written by $\tanh \alpha_\bot/\cosh \eta$.  
%Substituting the four momentum,   
%$ P^\mu =(m_T \cos y, P_T \cos \phi_p, P_T \sin \phi_p, m_T \sinh y)$ and 
% (\ref{eq-fourvel})
% for   
% (\ref{eq-particledist}), we obtain  
%The transverse spectrum of particles is then given by 
%\begin{equation}
%  \frac{dN_i}{d^2 P_Tdy} = \tau_{\text{SM}} \int\limits^\infty_{-\infty} 
%d\eta \int\limits^{2 \pi}_0 d \phi \int\limits^{r_{\text{max}}}_0 r \, dr 
%  \frac{P \cdot v}{(2\pi)^3}  G_i (P \cdot v )
%\end{equation}
%where $r=\tau_{\text{SM}} \sinh \eta_T$. 

Equation (\ref{eq:particledist}) can be evaluated analogous to 
(\ref{eq:partspec2}). We note that in the limit $P_T \to \infty$ Eqs.\ 
(\ref{eq:messpec},\ref{eq:barspec}) are equivalent to (\ref{eq:particledist})
if the same hypersurface and the same temperature and chemical potentials 
are used. This is an indication that recombination from a thermal parton
phase is the underlying microscopic picture of hadron production
in a statistical model. While
we will not elaborate on this
in more detail, we will quote some results of the statistical model 
for hadron ratios and compare with our calculation.

The geometric parameters are fixed to be 
$\tau_{\text{SM}}=7.66$ fm and $r_0=6.69$ fm for for most central 
collisions at RHIC in Ref.\ \cite{BroFlo:01}.
Particle ratios at mid rapidity in a boost-invariant model are not    
influenced by the expansion of the system \cite{BroFlo:01}, thus we can use 
the parameters which are determined by particle ratios from the entire phase 
space. We follow \cite{PBmMaReSt} and set $T_{\text{SM}}=177$ MeV, 
$\mu_B=29$ MeV, $\mu_S=10$ MeV and $\mu_I=-0.5$ MeV.

\subsection{Note on the parameters in our model.}

We want to give a brief summary of all the parameters for the parton phase.
Essentially we have three degrees of freedom for central collisions. Theses are
the energy loss given by $\epsilon_0 \langle L \rangle$, the slope of the
exponential part given by temperature $T$ and radial flow velocity $v_T$ 
and the normalization of the recombination spectrum by the volume
$\tau A_T$. In addition there are the parton fugacities.
After fixing $\langle L \rangle$, $T$ and $\tau$ to physical or at least
reasonable values, we retain $\epsilon_0$, $v_T$ and $\rho_0$ as true
parameters that were determined by fitting to the final data given by PHENIX
for the inclusive $\pi^0$ spectrum \cite{PHENIX:Adler03pi0auau}. This is in 
contrast to our previous study where the parameters of the parton spectrum 
were fixed by the peliminary charged hadron spectrum \cite{FMNB:03}.

The light quark fugacity was set to 1 in accordance with the measured $p/\pi^0$
ratio and the fugacities for antiquarks and strange quarks were obtained from
other ratios. The ratio of fugacities $\gamma_{\bar u}/\gamma_{u}=0.9
%=e^{-\mu_B/3T}
$ can be translated into a baryon chemical potential 
$\mu_B = 27$ MeV.
For other impact parameters, the simple geometric scaling of the volume and 
the number of collisions with $b$ and a reasonable ansatz for $\epsilon(b)$ 
describe the data up to $b=10$ fm. Only for very peripheral collision there
is the need to introduce the new parameter $\gamma(b)$.

%\emph{Anything for $v_2$ here?}

\section{Numerical Results}
\label{sec:results}

In this section we are going to discuss our numerical results
on hadron production.

\subsection{Hadron spectra}

\begin{figure}[t]
  \epsfig{file=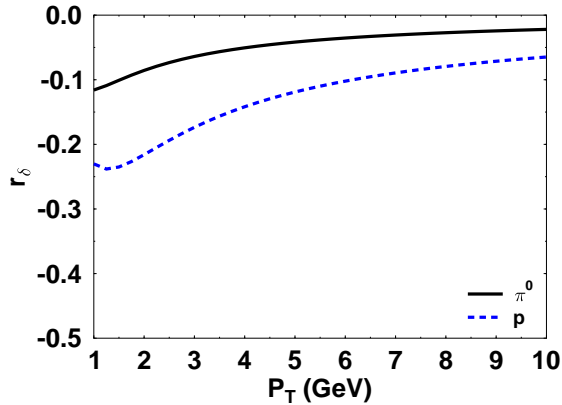,width=0.9\columnwidth}
  \caption{\label{fig:nwcomparison} Relative deviation $r_\delta$ of 
    the $\delta$ function approximation from calculations using wide wave 
    functions for neutral pions and protons.}
\end{figure}

In Fig. \ref{fig:spectra} we show our results for hadron production from
fragmentation and recombination for impact parameter $b=0$ in central Au+Au 
collision at $\sqrt{s}=200$ GeV.
We compare to available experimental data from the PHENIX and STAR 
collaborations at RHIC. 
The $\pi^0$ spectrum has been measured by PHENIX up to 10 GeV/$c$. The final
data were released very recently \cite{PHENIX:Adler03pi0auau} together
with final data for neutral pion production in $p+p$ collisions up to 
14 GeV/$c$ \cite{PHENIX:Adler03pi0pp}. 
Error bars show the total error in the $\pi^0$ yield.
All our calculations use the expressions (\ref{eq:res3},\ref{eq:protres}) 
with realistic light cone wave functions (\ref{eq:barlcwf}). 
For protons and neutral pions we also performed the calculations in the 
$\delta$-function approximation for the wave functions 
(\ref{eq:nwres},\ref{eq:nwres2}). 
Fig.\ \ref{fig:nwcomparison} shows the relative deviation 
$r_{\delta}=(dN - dN_{\delta})/dN$ of the $\pi^0$ and $p$ 
spectrum in $\delta$-function approximation from the recombination calculation 
using realistic wave functions. The deviation is less than 22\% for 
protons and less than 12\% for pions at small $P_T$. It becomes considerably 
smaller at larger $P_T$ since the violation of energy conservation is less
important there. This explains why calculations using 
(\ref{eq:nwres},\ref{eq:nwres2}) are often satisfactory.

Preliminary data on $\pi^+$ and $\pi^-$ production are only available up
to 2 GeV/$c$ \cite{Chujo:02PHENIX}, but we can expect that the global 
behavior of charged pions is similar to that of neutral pions. 
All data except for $\pi^0$, $p$ and $\bar p$ shown in Fig.\ 
\ref{fig:spectra} are preliminary.
Only statistical errors are given for all sets besided $\pi^0$.
In the pion spectra we clearly see the two $P_T$ domains of 
 hadron production. Above 4--5 GeV/$c$ the spectrum is dominated by 
fragmentation and follows a power law. Below 4 GeV/$c$ the spectrum is 
exponential and dominated by recombination from the thermal phase. In our
calculation the 
contribution from fragmentation is artificially cut off below 2 GeV/$c$ 
(corresponding to a lower cut-off of about 4 GeV/$c$ in the parton spectrum) 
because perturbative QCD loses its validity at low $P_T$. 

In the crossover region the yield is slightly underestimated, due to 
our simplified treatment not allowing for
recombination involving perturbative partons and mixed mechanisms 
\cite{GreKoLe:03}.
We also note that below 2 GeV/$c$ the calculated recombination spectrum bends 
down and underestimates the data. This effect is caused by neglecting
the large binding energy of pions in our recombination formalism, in which  
pions have an effective mass of $2 m_u \approx 520$ MeV compared to the 
true pion mass of 140 MeV. In addition, pions from secondary decays 
of hadronic resonances are an important contribution at very low $P_T$.

The same effect can be seen in the kaon spectra, where we again underestimate 
the yield for $P_T < 2$ GeV/$c$. 
The $K_s^0$ spectrum was measured by STAR up 6 GeV/$c$
\cite{Long:03STAR}. The preliminary $K^+$ and $K^-$ spectra up to 2 GeV/$c$ 
are available from PHENIX \cite{Chujo:02PHENIX}, and first results on 
$K^-$ from STAR up to 4 GeV/$c$ were shown recently \cite{Long:03STAR}. 
The $K_s^0$ data above 2 GeV/$c$ can
be described very well by our calculations. We note that the last four data 
points of the preliminary STAR data on $K^-$ follow a different systematics 
than the rest of the points and also seem to deviate from the $K_s^0$ data. 
This could indicate a failure of proper particle identification in this
momentum range.

\begin{figure}[t]
  \epsfig{file=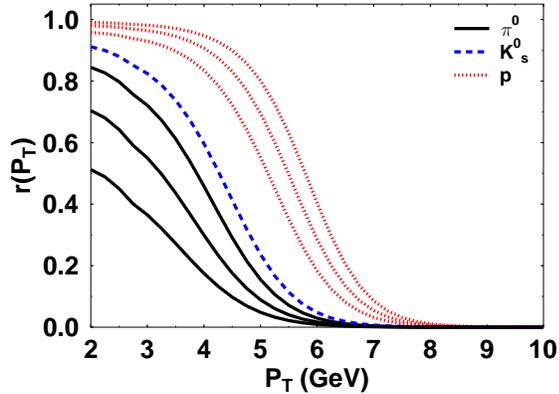,width=0.9\columnwidth}
  \caption{\label{fig:RFratios} The ratio $r(P_T)=R/(R+F)$ of recombined 
   hadrons to the sum of recombination and fragmentation for $\pi^0$ (solid), 
   $K_s^0$ (dashed) and $p$ (dotted lines). For protons and pions
   different impact parameters $b=0$, 7.5 and 12 fm (from top to bottom)
   are shown. $K_s^0$ is for $b=0$ fm only.}
\end{figure}

Protons and antiprotons have been identified by the PHENIX collaboration
up to 4.5 GeV/$c$ and the final data was published very recently
\cite{PHENIX:03ppbar}. In contrast to the Goldstone bosons
considered before, the mass of $p$ and $\bar p$ in our constituent quark
picture is closer to the physical mass of 938 MeV and secondary protons 
and antiprotons are less abundant.
Hence our calculation provides a satisfactory description of the
spectra even between 1 and 2 GeV/$c$. This is true for all hadrons 
that are not Goldstone bosons. The crossover between the recombination
and the fragmentation process is shifted upward for $p$ and $\bar p$
compared with pions, to around 6 GeV/$c$.  We alert the reader to the 
apparent suppression of the fragmentation process below 6 GeV/$c$ 
compared with recombination. This is much more prominent here than 
in the case of pions.  In Fig.\ \ref{fig:RFratios} we show the
ratio $r(P_T)$ of recombined hadrons to the full calculation, defined in
Eq. (\ref{eq:recoweight}).
The shift of the crossover to higher $P_T$ from pions over kaons
to protons is obvious. The 50\% mark changes from 4 to 4.5 to 6 GeV/$c$.
%This is mainly due to the decreasing probability for the heavier
%hadrons to be formed by 
%fragmentation. 

Here we need to emphasize an important point regarding the 
perturbative calculation.  The fragmentation functions are a
non-perturbative input to these calculations, and are derived from 
other experiments. Most data about fragmentation functions are from 
$e^+ e^-$ annihilation experiments which do not allow to distinguish 
between quark and antiquark fragmentation. Quarks have to be created in 
pairs, so that only fragmentation functions like $D_{u+\bar u\to p+\bar p}$ 
can be deduced. Additional input from semi-exclusive reactions helps to
separate the contributions, but fragmentation functions still require 
improvement for applications in hadron interactions. 
The KKP parametrization seems to work well for $\pi^0$ production 
at RHIC \cite{PHENIX:Adler03pi0pp} but we anticipate more problems
for other hadrons, in particular for protons and antiprotons. This 
suggests a considerably larger theoretical uncertainty for the results 
from fragmentation of all hadrons other than pions. For this reason,
and due to the lack of appropriate fragmentation 
functions,
we do not show the fragmentation contribution for $\Phi$, $\Xi$ 
and $\Omega$ in our calculations. 

Our results for $\Lambda+\bar\Lambda$ include an equally large
contribution from $\Sigma^0$ and $\bar\Sigma^0$. The preliminary STAR data
is taken from \cite{Long:03STAR}. For the multistrange hadron spectra
$\Xi^-+\Xi^+$, $\Omega+\bar\Omega$ and $\Phi$ -- which is supposed to have 
a pure $s\bar s$ valence structure -- we present only recombination spectra. 
The strange hadron yields determine our value of the strange quark 
fugacity $\gamma_s = 0.8$.
Preliminary STAR data is available on $\Xi^-+\Xi^+$ \cite{Long:03STAR} 
and $\Omega+\bar\Omega$ (minimum bias) \cite{Suire:02STAR}. We also 
show a calculation for an impact parameter of 10 fm which agrees well with 
the minimum bias $\Omega+\bar \Omega$ data. No spectra for $\Phi$ mesons 
at RHIC have been published so far.

\begin{figure}[b]
  \epsfig{file=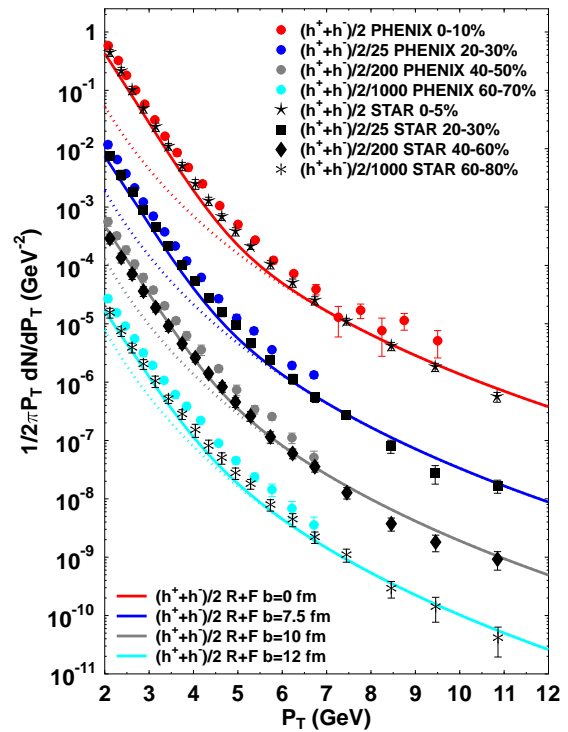,width=0.9\columnwidth}
  \caption{\label{fig:chhad} The spectrum of charged hadrons $(h^+ =h^-)/2$
  for four different impact parameters 0, 7.5 (divided by 25), 10 (/200) and 
  12 fm (/1000) [from top to bottom] in comparison with data from STAR and 
  PHENIX in different centrality bins. Contributions 
  from fragmentation only (dotted) and the sum of recombination and 
  fragmentation (R+F, solid lines) are shown.}
\end{figure}

The charged hadron spectrum $(h^+ +h^-)/2$ is shown separately in Fig.\
\ref{fig:chhad}:  pions, protons, antiprotons and kaons are
taken in into account.
This spectrum, including its impact parameter dependence, will be
discussed in detail in the section on centrality dependence.

In summary, our calculations using one fixed parameter set are consistent with 
all the currently existing data from RHIC and 
allow us to make predictions for
future measurements.

\subsection{Hadron ratios}

\begin{figure*}
  \epsfig{file=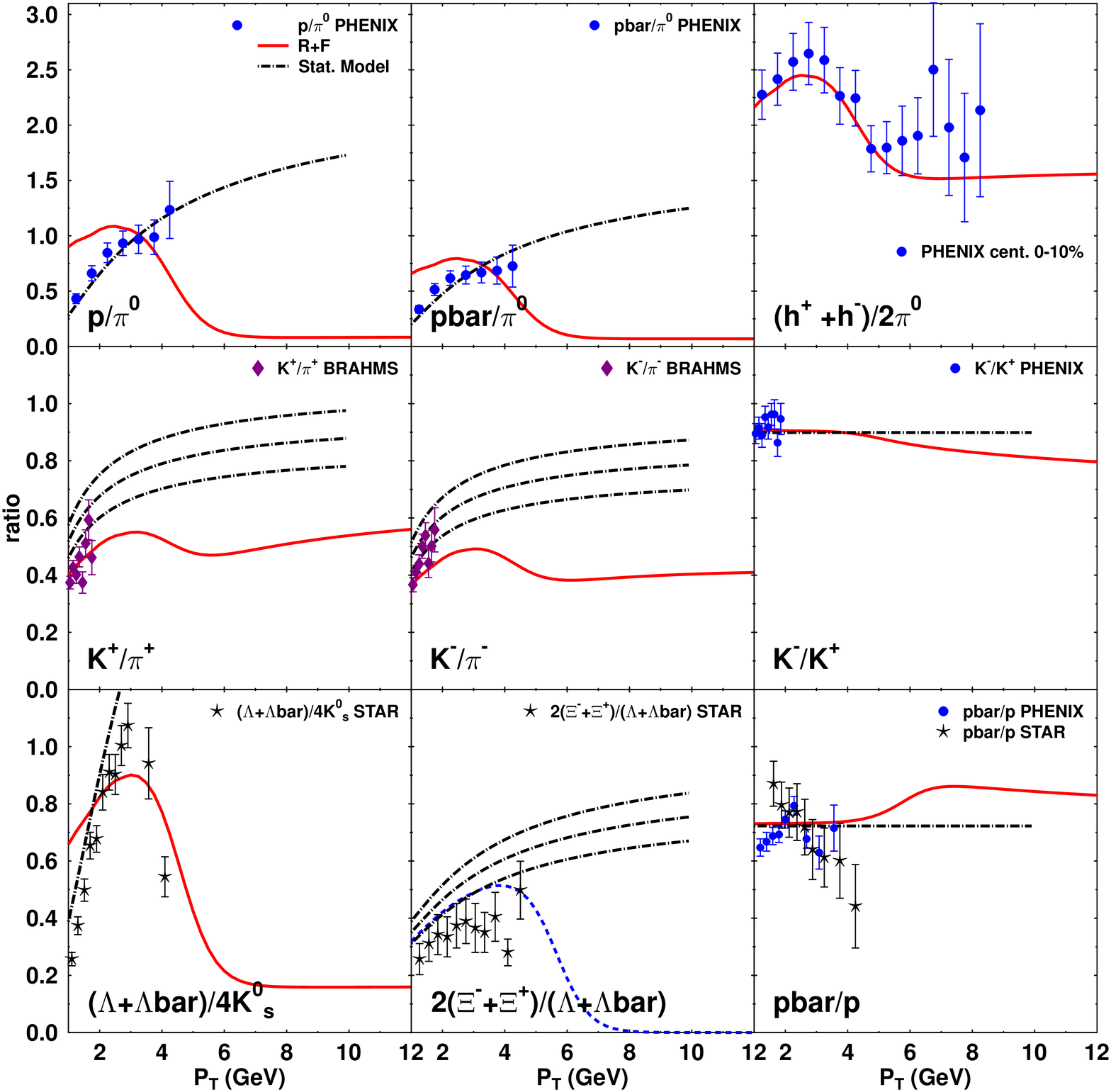,width=0.9\textwidth}
  \caption{\label{fig:ratios} Hadron ratios $p/\pi^0$, $\bar p/\pi^0$,
    $\bar p/p$, $K^+/\pi^+$, $K^-/\pi^-$, $K^-/K^+$, $(\Lambda+\bar\Lambda)/
    4 K_s^0$, $2(\Xi^-+\Xi^+)/(\Lambda+\bar\Lambda)$ and $2 \pi^0/(h^+ + h^-)$
    as functions of transverse momentum $P_T$. The calculation for $\Xi$ 
    baryons only takes into account recombination. We show data from STAR,
    PHENIX and BRAHMS and results from the statistical model (dash dotted 
    lines). Where several curves for the statistical model are shown these
    are for different strangeness fugacities 1.0, 0.9 and 0.8 (from top to 
    bottom).}
\end{figure*}

In Fig.\ \ref{fig:ratios} we show hadron ratios. Only statistical errors 
are shown. The systematic errors can 
be quite large -- we refer the reader to the cited experimental publications
for further details.

One of the main motivations at the onset of our investigation was
to find an explanation for the 
surprisingly large proton over pion 
ratios that are of order one above 1.5 GeV/$c$. 
The data on the $p/\pi^0$ and $\bar p/\pi^0$ ratios stem from the PHENIX 
collaboration \cite{PHENIX:03ppbar}. 
We have already shown in a previous publication 
\cite{FMNB:03} that recombination naturally 
provides a $p/\pi^0$ of order one for hadron transverse momenta up to 4 GeV. 
In addition we predict that a sharp drop
beyond 4 GeV/$c$ should be seen when the fragmentation process takes over.
The value predicted for the ratio in the fragmentation domain is about 0.1.
At small transverse momenta, the  
statistical model describes the data well but continues
to rise beyond 4 GeV/$c$.

The $K^+/\pi^+$ and $K^-/\pi^-$ ratios have been measured by BRAHMS
\cite{Lee:03BRAHMS}, and the $K^-/K^+$ ratio is compared to data from PHENIX
\cite{Chujo:02PHENIX}. For the statistical model we probe an additional
strangeness fugacity and provide curves for values of 1.0, 0.9 and 0.8. For
the recombination part $\gamma_s =0.8$ was used as discussed above.
The interesting feature of the preliminary $\bar p/p$ data from STAR
\cite{Kunde:02STAR} and PHENIX \cite{Chujo:02PHENIX} is that PHENIX
implies a flat $\bar p/p$ ratio up to 4 GeV/$c$, while STAR sees a decrease
from 2 GeV/$c$ on. (This may be related to the apparent surplus of $K^-$
measured by STAR in the same momentum range, indicating a possible failure 
of charged particle identification beyond 2 GeV/$c$.)  Our calculation 
predicts a flat ratio in this transverse momentum range in agreement 
with the PHENIX data. A linear sum of statistical and systematic error 
is shown for the STAR data. 

Preliminary results for the ratios $(\Lambda+\bar\Lambda)/4 K_s^0$ and 
$2(\Xi^-+\Xi^+)/(\Lambda+\bar\Lambda)$ were also presented from the STAR
collaboration \cite{Long:03STAR}. Our calculations (for $\Xi$ recombination 
only) are in rather good agreement. Recombination predicts a large peak 
in the $\Lambda$/kaon ratio and a sharp decrease beyond 4 GeV/$c$ similar to 
the $p/\pi^0$ ratio. First indications of such a sharp transition can be 
seen in the STAR data. This observation supports the recombination picture 
including a transition to the fragmentation regime beyond 4 GeV/$c$.

For completeness we also show the ratio of charged hadrons to neutral
pions reported by PHENIX \cite{PHENIX:03ppbar} in
comparison with our results. Our calculation 
agrees very well with the PHENIX data in the recombination region but 
slightly underestimates the ratio in the fragmentation region. 

In summary, our calculations are in good agreement with the available RHIC 
data. The predicted decrease in the proton/pion ratio, if confirmed, and 
first observations of a similar drop in the $\Lambda$/kaon ratio will be 
strong arguments in favor of the recombination+fragmentation picture.

\subsection{Centrality dependence}

\begin{figure}[t]
  \epsfig{file=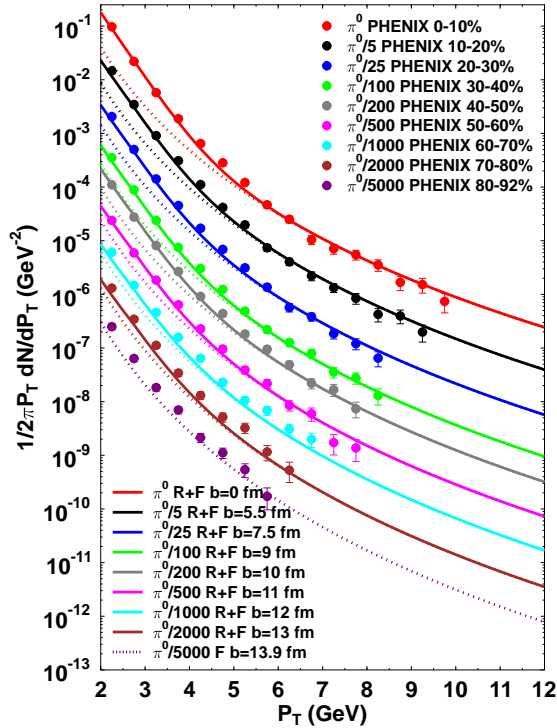,width=0.9\columnwidth}
  \caption{\label{fig:pi0spec} The spectrum of neutral pions for
  impact parameters 0, 5.5 (divided by 5), 7.5 (/25), 9 (/100), 10 (/200),
  11 (/500), 12 (/1000), 13 (/2000) and 13.9 fm (/5000) [from top to bottom]
  in comparison with data from PHENIX in different centrality bins. 
  Contributions from fragmentation only (dotted) and the sum of recombination 
  and fragmentation (R+F, solid lines) are shown. 13.9 fm calculation is 
  fragmentation (F) only.}
\end{figure}

\begin{figure}[t]
  \epsfig{file=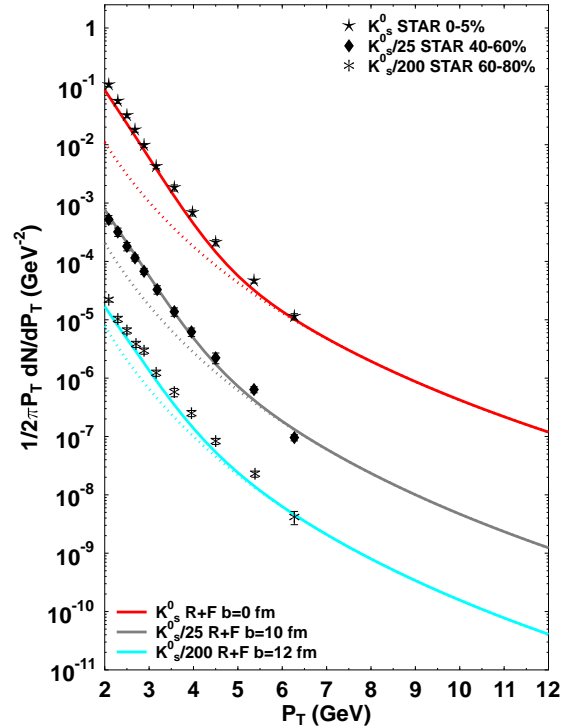,width=0.9\columnwidth}
  \caption{\label{fig:k0sspec} The spectrum of neutral kaons for
  impact parameters 0, 10 (/25) and 12 fm (/200) [from top to bottom] in 
  comparison with data from STAR in different centrality bins. Contributions 
  from fragmentation only (dotted) and the sum of recombination and 
  fragmentation (R+F, solid lines) are shown.}
\end{figure}

Figs.\ \ref{fig:pi0spec} and \ref{fig:k0sspec} show the centrality
dependence of the 
$\pi^0$ and $K^0_s$ spectra.
Final results in various 
centrality bins for $\pi^0$ have been published by 
PHENIX \cite{PHENIX:Adler03pi0auau}. 
The $K^0_s$ data are preliminary results
from STAR \cite{Long:03STAR}. The impact parameter dependence of the charged
hadron spectrum $(h^++h^-)/2$ was already shown in Fig.\ \ref{fig:chhad}
with data from STAR \cite{STAR:03chhad} and PHENIX \cite{Velk:03PHENIX}.

The impact parameter dependence of the parameters in our calculation 
was fixed by a fit to the $\pi^0$ 
data but it is consistent with the kaon and charged hadron data. We notice
that with increasing impact parameter the hadrons from fragmentation come 
ever closer to the data points below the crossover point. Thus 
fragmentation becomes more and more important in peripheral collisions
in accordance with our expectations. For the most peripheral bin in $\pi^0$,
with an impact parameter of 13.9 fm we refrained from extracting a
recombination contribution. In principle the data can be explained by
fragmentation alone down to 2 GeV/$c$. We give a recombination contribution
for $b=13$ fm although its contribution 
might be disputable already in this centrality bin.

\begin{figure}[t]
  \epsfig{file=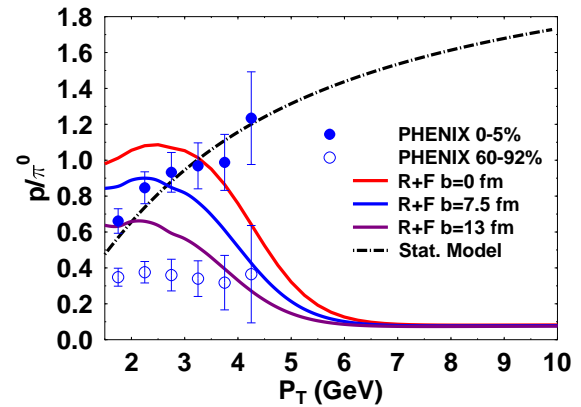,width=0.9\columnwidth}
  \caption{\label{fig:protpiratio} The ratio $p/\pi^0$ for three different
   impact parameters 0, 7.5 and 13 fm (solid lines, top to bottom)
   compared to the statistical model (dashed dotted line) and PHENIX data.}
\end{figure}

Fig.\ \ref{fig:RFratios} shows the ratio of $r(P_T)$ for three
different impact parameters (0, 7.5 and 12 fm) for protons and pions. 
The systematics confirms that the crossover point shifts to 
smaller $P_T$ for increasing $b$ and that the weight of fragmentation 
becomes more and more important. In Fig.\ \ref{fig:protpiratio} we display
the impact parameter dependence of the $p/\pi^0$ ratio. As expected the 
proton/pion ratio is decreasing with increasing $b$ in the recombination
region and is unaltered where fragmentation is dominating.

It is an interesting question at which impact parameter the recombination
mechanism becomes negligible for transverse momenta above 2 GeV/$c$. 
This will happen at smaller impact parameter for pions than for protons. 
Related to this issue is the question whether recombination contributes to
hadron production at central rapidity in $p+A$ or $d+A$ reactions, 
where the produced matter is less dense.  We still expect recombination 
to be an important mechanism at low $P_T$.  However, there will probably 
be no chemical and thermal equilibration in the parton spectrum, and much 
less flow.  The $b$ dependence in Au+Au collisions is not a good basis 
for extrapolation since flow and equilibration seem to diminish only in 
very peripheral collisions. 
That makes it difficult to give quantitative predictions without data.
However we would expect that no recombination effects in pion production 
are visible above 2 GeV/$c$ in $d+$Au collisions.

\subsection{Nuclear modification factors}

The nuclear modification factor is defined as the ratio of the hadron yield
in Au+Au collisions to the one in $p+p$ scaled with the number of collisions
\begin{equation}
  R_{AA} = \frac{d^2N_{\text{Au+Au}}(b)/dP_T^2}{N_{\text{coll}}(b) \, 
  d^2N_{{p+p}}/dP_T^2}.
\end{equation}
Similarly one can consider scaled ratios of different centrality bins
like central to peripheral
\begin{equation}
  R_{CP} = \frac{N_{\text{coll}}(b)\, d^2N_{\text{Au+Au}}(0)/dP_T^2}
  {N_{\text{coll}}(0)\, d^2N_{\text{Au+Au}}(b)/dP_T^2}.
\end{equation}

\begin{figure}[t]
  \epsfig{file=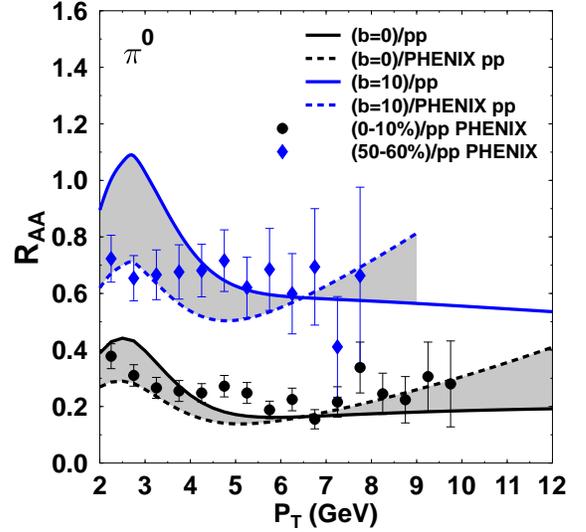,width=0.9\columnwidth}
  \caption{\label{fig:raapi0} Nuclear modification factor $R_{AA}$ for impact 
   parameters 0 (bottom) and 10 fm (top) . Normalization by $p+p$ is via our 
   own calculation (solid) or via PHENIX $p+p$ results (dashed lines). 
   Data on $R_{AA}$ are from the PHENIX collaboration with point-to-point 
   errors only.}
\end{figure}

In Fig.\ \ref{fig:raapi0} we show the nuclear modification factor for neutral
pions for two impact parameters, $b=0$ and $b=10$ fm. We provide  
ratios taken both with our own $p+p$ calculation and with the $p+p$ results
from PHENIX \cite{PHENIX:Adler03pi0pp} and compare to final data from
PHENIX \cite{PHENIX:Adler03pi0auau}. We notice that there is an apparent
uncertainty in the perturbative calculation which makes the two curves
using our own $p+p$ calculation and the PHENIX $p+p$ results deviate. However,
both curves are consistent with the data for central collisions. The problem
is amplified for peripheral collisions, where the curve using our $p+p$
calculation overestimates $R_{AA}$ below 4 GeV/$c$. The spread between
both curves can be interpreted as a typical error to be expected in a
lowest-order perturbative calculation.
The nuclear modification factor for pions shows the strong jet quenching
effect that suppresses the pion yield by about a factor of 5 for the highest
$P_T$ bins in central collision. Recombination predicts a slight increase 
below 4 GeV/$c$ which can be observed in the data. However, recombination is 
not able to compensate the loss of pions through jet quenching at high $P_T$.
In peripheral collisions jet quenching effects are much weaker.

\begin{figure}[t]
  \epsfig{file=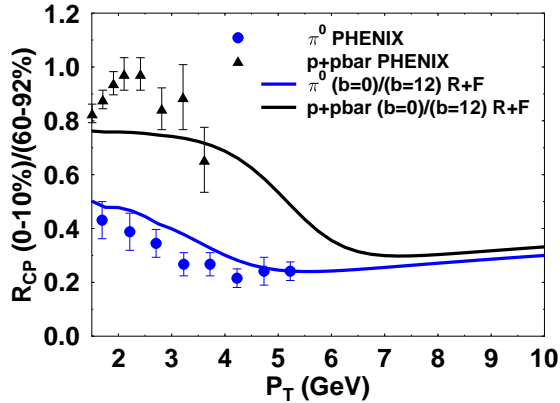,width=0.9\columnwidth}
  \caption{\label{fig:rcppiprot} $R_{\text{CP}}$ for neutral pions (bottom) 
   and protons (top) given by the ratio of particle yields at impact 
   parameters 0 and 12 fm compared to data from PHENIX.}
\end{figure}

Fig.\ \ref{fig:rcppiprot} displays the scaled ratio $R_{\rm CP}$ for neutral 
pions and protons. The ratio of impact parameters 0 and 12 fm is used and 
compared to data from the PHENIX collaboration \cite{Velk:03PHENIX}. 
The data for protons shows
$R_{\rm CP}$ to be between 0.8 and 1 below 4 GeV/$c$, which is quite surprising,
considering the strong suppression suffered by the pions in that momentum
domain.
As already noticed above, recombination is more effective for protons than 
for pions. Therefore  our calculation for the protons 
yields a similarly value of 0.8 as observed by the experiment.
This implies
that protons from recombination make up for the loss suffered from jet 
quenching 
at intermediate transverse momenta. Our calculations predicts sharp drops in 
$R_{\rm CP}$ and $R_{AA}$ for protons and antiprotons beyond 4 GeV/$c$ where 
fragmentation with jet quenching start to dominate.

\begin{figure}[t]
  \epsfig{file=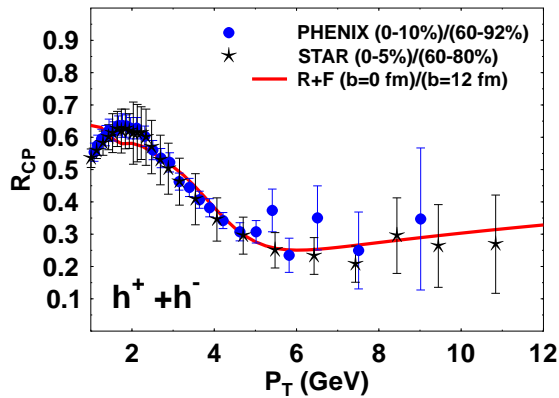,width=0.9\columnwidth}
  \caption{\label{fig:raachhad} $R_{\text{CP}}$ for charged hadrons given by 
   the ratio of particle yields at impact parameters 0 and 12 fm compared 
   to data from STAR and PHENIX.}
\end{figure}

In Fig.\ \ref{fig:raachhad} we give $R_{\rm CP}$ for charged hadrons compared
to data from STAR \cite{Klay:02STAR} and PHENIX \cite{Miod:02PHENIX}.
The contribution from recombination below 4 GeV/$c$ leads to a value
of about 0.6 which is between the values for protons and pions.
The observed steep drop to the value attributed to jet quenching is well 
described by our theory.

\begin{figure}[t]
  \epsfig{file=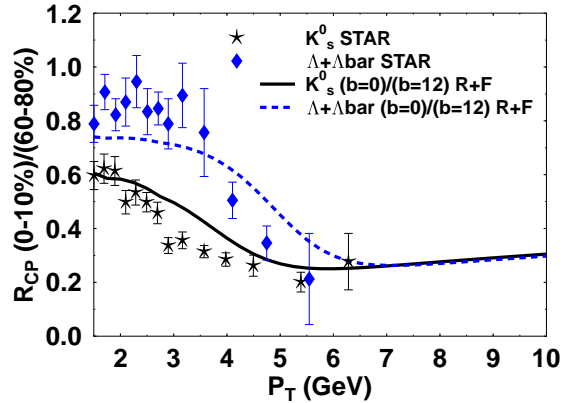,width=0.9\columnwidth}
  \caption{\label{fig:rcpk0sllbar} 
   $R_{\text{CP}}$ for $K^0_s$ (bottom) and $\Lambda+\bar\Lambda$ (top)
   given by the ratio of particle yields at impact parameters 0 and 12 fm 
   compared to data from STAR.}
\end{figure}

Finally in Fig.\ \ref{fig:rcpk0sllbar} we compiled $R_{\rm CP}$ for $K_s^0$
and $\Lambda+\bar\Lambda$ together with data from STAR \cite{Long:03STAR}.
The different behavior of mesons and baryons is again impressively confirmed.
For the first time experimental data indicate that a steep decrease in
$R_{\text{CP}}$ for baryons will occur beyond 4 GeV/$c$.
The data on $\Lambda+\bar\Lambda$ suggest a drop to the perturbative value 
even sharper than what our results show. This 
could be due to too less $\Lambda$ baryons from fragmentation using this
particular set of fragmentation functions \cite{dFSV:97}.

\subsection{Results on elliptic flow}

\begin{figure}[t]
  \epsfig{file=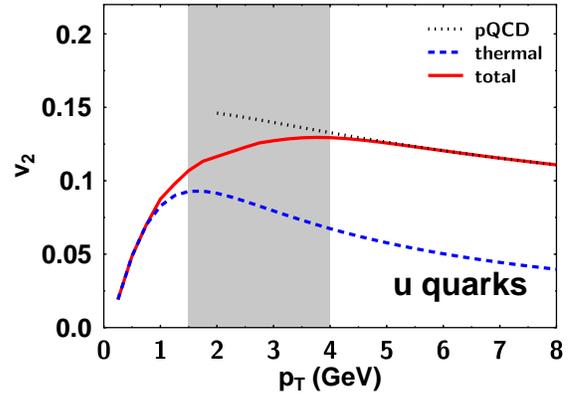,width=0.9\columnwidth}
  \caption{\label{fig:v2parton} 
   Elliptic flow $v_2$ in the parton phase as a function of transverse 
   momentum $p_T$. The flow in the thermal phase (dashed) 
   and a pQCD calculation (dotted line) are shown. The solid line 
   interpolates between the two domains. The shaded region shows the
   region where the interpolation takes place.}
\end{figure}

\begin{figure}[t]
 \epsfig{file=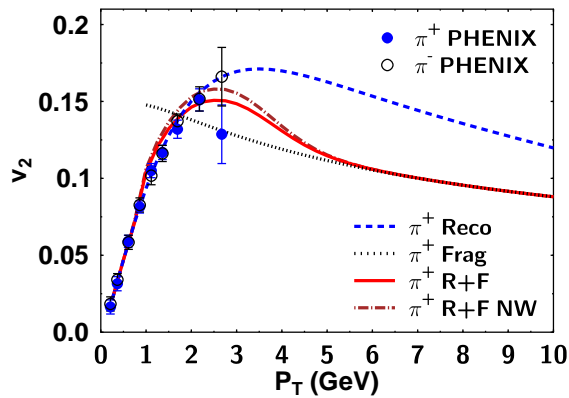,width=0.9\columnwidth}
    \caption{\label{fig:v2pion} $v_2$ for positively charged pions. We show 
   recombination only (dashed), fragmentation only (dotted) and the full 
   calculation (solid line).
   The result of a calculation in the $\delta$-function approximation (NW) for
   the wave function is also shown (dash-dotted line). Data are taken from the 
   PHENIX collaboration \cite{PHENIX:v2}. }
\end{figure}

\begin{figure}[t]
  \epsfig{file=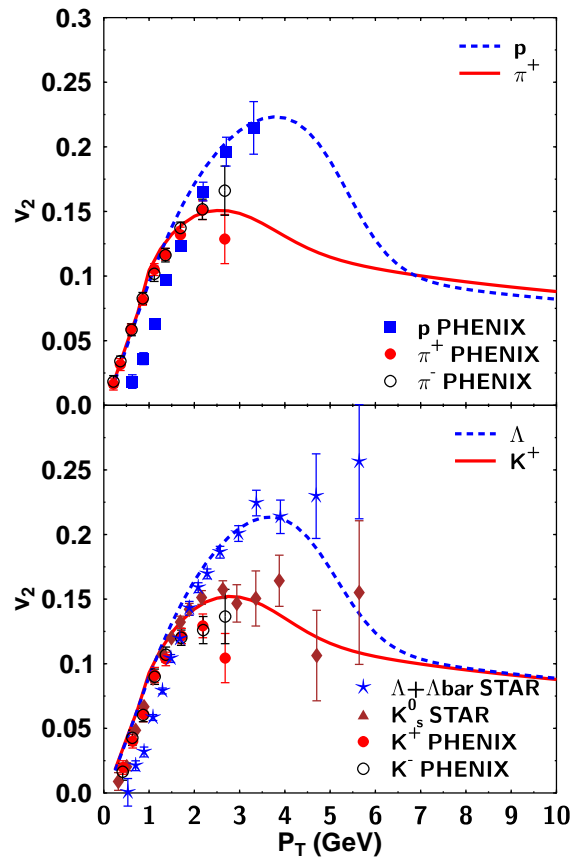,width=0.9\columnwidth}
  \caption{\label{fig:v2} 
   Upper panel: anisotropic flow for $p$ and $\pi^+$ compared to
   PHENIX data \cite{PHENIX:v2}. Lower panel: $v_2$ for $K^+$ and 
   $\Lambda+\bar\Lambda$ compared to preliminary STAR data 
   ($\Lambda+\bar\Lambda$, $K^0_s$) 
   \cite{Snellings:03} and PHENIX data ($K^+$, $K^-$) \cite{PHENIX:v2}.}
\end{figure}

\begin{figure}[t]
  \epsfig{file=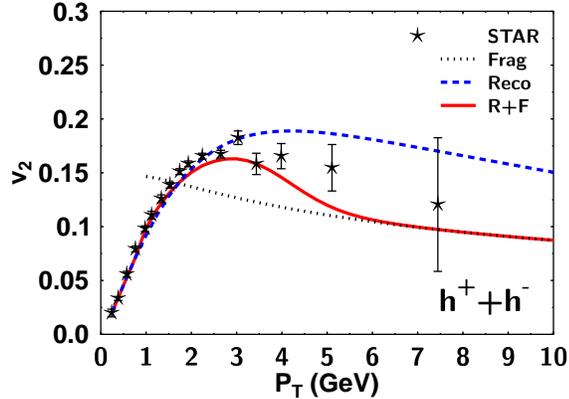,width=0.9\columnwidth}
  \caption{\label{fig:v2had} 
   $v_2$ for charged hadrons. Again we show the contributions from
   different mechanisms as in Fig.\ \ref{fig:v2pion}. 
   Data are preliminary and taken from the STAR collaboration.}
\end{figure}

\begin{figure}[t]
  \epsfig{file=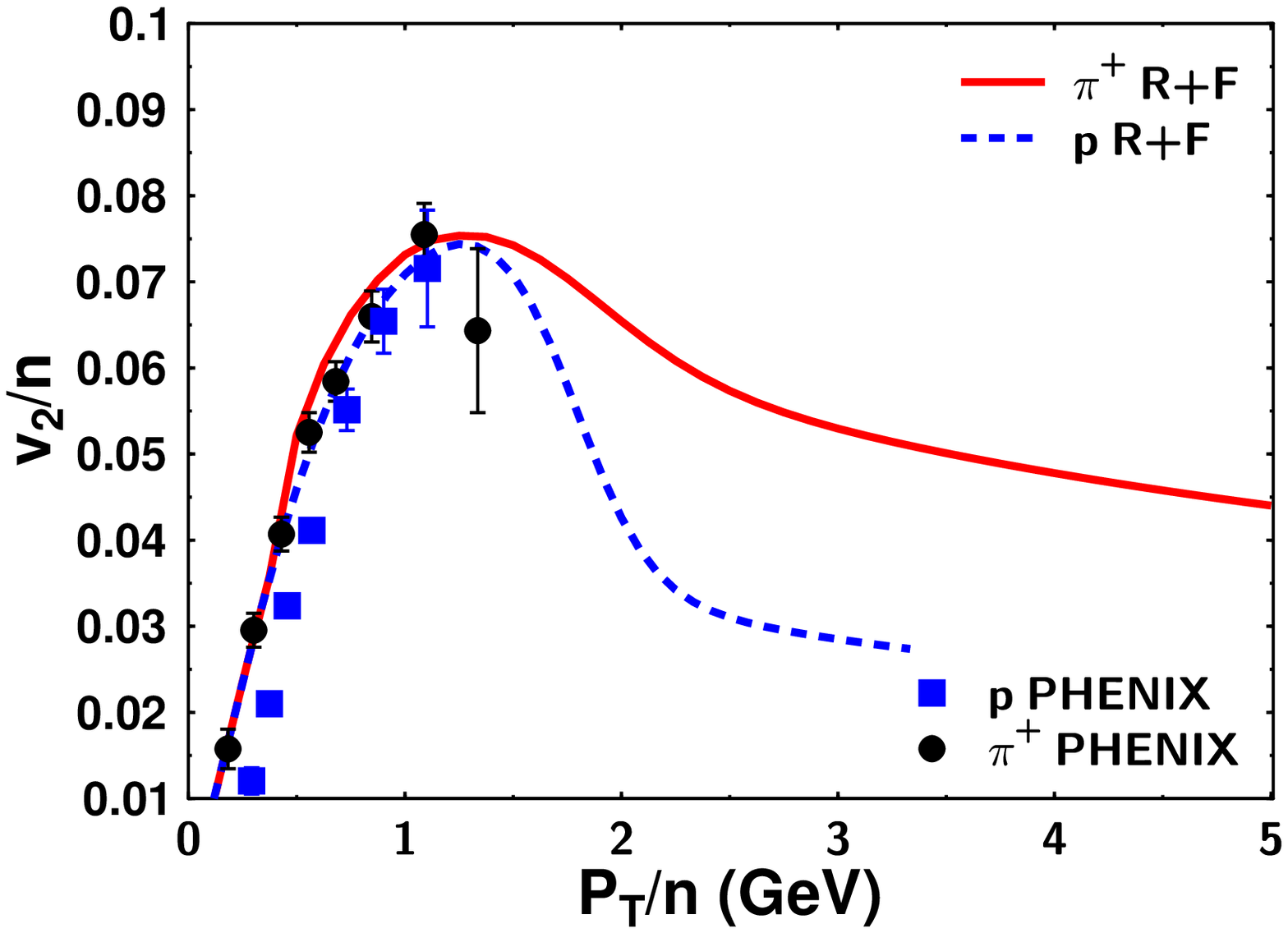,width=0.9\columnwidth}
  \caption{\label{fig:v2scaled} 
  The anisotropy $v_2/n$ for pions (bottom) and protons (top) 
  as a function of transverse momentum $p_T/n$ using the scaling law
  (\ref{eq:v2scaling}) with $n=2$ for pions and $n=3$ for protons. 
  Data points are pions and protons from PHENIX using the same scaling law.}
\end{figure}

Fig.\ \ref{fig:v2parton} shows the elliptic flow $v_2(p_T)$ of $u$ quarks
before hadronization. The contributions from jet quenching and
from anisotropic flow in the thermal phase are shown separately as well. 
Due to the very different
behavior in the two domains, $v_2$ is more sensitive to mechanisms which
interpolate between the perturbative domain and the soft domain. 
The range of this theoretical uncertainty is highlighted by the 
shaded region in 
Fig.\ \ref{fig:v2parton}.

In Fig.\ \ref{fig:v2pion} we provide results for $\pi^+$. The
contributions from fragmentation and recombination are shown separately. 
The full
calculation interpolates between these two curves in the interval between
2 and 4 GeV/$c$. We also compare the result of a calculation using the
$\delta$-function approximation for the wave functions. As expected the
deviations are small. Our calculations agree with PHENIX data on $v_2$ of 
$\pi^+$ and $\pi^-$ \cite{PHENIX:v2}.

Our results on the particle dependence of elliptic flow are summarized in
Fig.\ \ref{fig:v2}. One can see the different behavior of mesons and baryons
by comparing protons with pions and kaons with $\Lambda$s. 
$v_2$ for baryons saturates at a higher value than for mesons in the 
recombination domain. At higher $P_T$, when fragmentation takes over, 
the results rapidly approach each other. In our calculation, where we do not
take into account the binding energies, we cannot resolve the splitting
between protons and mesons coming from the mass difference. Nevertheless
the agreement with data from PHENIX \cite{PHENIX:v2} and preliminary data 
from STAR \cite{Snellings:03} is good. In Fig.\ \ref{fig:v2had} $v_2$ for
charged hadrons is compared with preliminary STAR data \cite{Tang:CIPANP}. 
This is interesting since for charged hadrons the measurements extend 
up to 7 GeV/$c$ and constrain $v_2$ in the pQCD domain. Note that in this
case $v_2$ was extracted by STAR from 4 particle correlations. This is 
supposed to reduce non-flow effects to $v_2$ \cite{BoDinOll:01} in comparison
with the usual reaction plane analysis.

In Fig.\ \ref{fig:v2scaled} we test the scaling law from Eq.\ 
(\ref{eq:v2scaling}) for protons and pions. 
%Here one can see, that the flow in the thermal parton phase was chosen
%to describe the PHENIX data on the elliptic flow of pions \cite{Esumi:02}. 
Protons and pions follow one universal curve below 1.5 GeV/$c$, which is very
similar to the flow of thermal partons given in Fig.\ \ref{fig:v2parton}. 
Beyond 1.5 GeV/$c$ we predict a transition to the values given by pQCD. The 
scaling law is no longer valid in that domain.

We would like to emphasize that
we expect modifications from other hadronization mechanisms in the
region where we interpolate between the recombination and the fragmentation
dominated domains. These could be quite important in the case of $v_2$ and
alter the results in the interpolation region, e.g.\ they could smoothen
the transition between both domains. Interactions in the hadronic
phase could alter $v_2$ further.

\section{Conclusions}

In this work we have presented extensive evidence that recombination is 
the dominant hadronization mechanism for central Au+Au collisions at RHIC 
up to about 4 GeV/$c$ for pions and 6 GeV/$c$ for protons. 
We have described a covariant framework that permits the calculation 
of recombination from a dense thermal parton phase using light-cone 
wave functions for the produced hadrons. This formalism is adequate for 
momenta much larger than the non-perturbative scales involved. At lower 
energies, energy and entropy conservation pose a serious problem, the
solution of which requires a dynamical, rather than purely kinematic, 
treatment of 
the recombination process. We have found that, for practical purposes,
hadron spectra are well described down to transverse momenta of 2 GeV/$c$ 
for Goldstone bosons ($\pi,K$) and 1 GeV/$c$ for other hadrons.

Recombination competes with fragmentation from perturbatively scattered 
partons.  The large energy loss of these partons leads to sizable quenching 
factors, which reduce the fragmentation contribution and cause it to be 
buried under soft physics at scales which one would not generally attribute 
to soft physics. However, 4 GeV/$c$ in the pion and 6 GeV/$c$ in the proton 
spectrum correspond to a transverse momentum of only 2 GeV/$c$ on average 
for the coalescing partons.

The interplay of recombination and fragmentation leads to interesting 
effects in particle ratios and nuclear modification factors. 
The proton/pion ratio is naturally around one in the recombination regime.
For protons the unquenching effect by recombination below 4 GeV/$c$ is so 
strong that essentially no nuclear suppression can be observed at all
in this momentum range. For the proton/pion ratio and suppression factors 
$R_{AA}$ and $R_{\rm CP}$ we expect a sharp drop beyond 4 GeV/$c$ indicating 
the beginning of the perturbative regime.

For the azimuthal asymmetry $v_2$, our calculations describe the data well.
The different behavior of baryons and mesons is above 1 GeV/$c$ can be
explained. The scaling law (\ref{eq:v2scaling}), derived
from the recombination formalism, is consistent with data up to 1.5 GeV/$c$. 
We predict a violation of this scaling law at higher values, coming from 
perturbative QCD.

In this publication we have only considered single hadron production and 
neglected correlations in the hadron emission pattern. The yield of secondary 
hadrons, when triggering on a leading hadron, is a promising quantity to 
provide more information about the underlying hadronization mechanism.

With fragmentation and energy loss alone, no consistent 
explanation involving \emph{all} hadron species can be given. In contrast
we are able to describe most available RHIC data on spectra, ratios,
nuclear suppression and elliptic flow of hadrons, including their impact 
parameter dependence, for transverse momenta above 1--2 GeV/$c$ -- 
for $v_2$ even down to very low $P_T$ -- consistently with a very small
number of globally adjusted parameters. As input for the recombination 
process we use a dense phase of partons with temperature $T=175$ MeV and 
radial flow velocity $v_T=0.55c$ at hadronization time 5 fm. All RHIC 
data shown in this work are consistent with the existence of such a phase.

\begin{acknowledgments}
This work was supported by RIKEN, Brookhaven National Laboratory, 
DOE grants DE-FG02-96ER40945 and DE-AC02-98CH10886, and by the 
Alexander von Humboldt Foundation. We thank M.\ Stratmann for providing
a code for $\Lambda$ fragmentation. We are grateful to D.\ d'Enterria,
S.\ D.\ Ellis, D.\ Hardtke, U.\ Heinz, P. Jacobs, Z.\ W.\ Lin, D.\ Molnar, 
J. Velkovska and X.\ N.\ Wang for 
useful discussions.
\end{acknowledgments}

\end{document}